\documentclass[acmtog,screen,nonacm]{acmart}

\AtBeginDocument{%
  }

\setcopyright{none}
\acmDOI{}
\usepackage{amsmath}
\usepackage{xspace}
\usepackage{graphicx}
\usepackage{wrapfig}
\usepackage{subcaption}
\usepackage{enumitem}

\usepackage{multirow}
\usepackage{makecell}

\newcommand{\ti}{$\times$}

\newcommand{\amulet}{\textit{Amulet}\xspace}
\begin{document}

\newcommand{\mytitle}{Amulet: Frame Extrapolation Through Sparse Layered Scene Representation and Adaptive Shading}
\title{\mytitle}

\newcommand{\affiliationVISUS}{%
\affiliation{%
  \institution{University of Stuttgart}
  \city{Stuttgart}
  \country{Germany}}
}

\author{Sebastian Künzel}
\orcid{0009-0001-0799-4293}
\authornote{These authors contributed equally to this work.}
\authornote{Corresponding author: sebastian.kuenzel@visus.uni-stuttgart.de.}
\affiliationVISUS{}
\email{sebastian.kuenzel@visus.uni-stuttgart.de}

\author{Fabian Schmierer}
\orcid{0009-0000-0117-326X}
\authornotemark[1]
\affiliationVISUS{}
\email{fabian.schmierer@visus.uni-stuttgart.de}

\author{Sergej Geringer}
\orcid{0000-0001-5147-9785}
\affiliationVISUS{}
\email{sergej.geringer@visus.uni-stuttgart.de}

\author{Guido Reina}
\orcid{0000-0003-4127-1897}
\affiliationVISUS{}
\email{guido.reina@visus.uni-stuttgart.de}

\author{Daniel Weiskopf}
\orcid{0000-0003-1174-1026}
\affiliationVISUS{}
\email{daniel.weiskopf@visus.uni-stuttgart.de}

\author{Dieter Schmalstieg}
\orcid{0000-0003-2813-2235}
\affiliationVISUS{}
\email{dieter.schmalstieg@visus.uni-stuttgart.de}

\begin{abstract}
We introduce \amulet, a rendering method that transforms a scene into a sparse, tiled and layered intermediate scene representation (cache) for high-frequency frame extrapolation. In contrast to reprojection-based techniques, \amulet explicitly rasterizes and stores potentially visible geometry in its layered image-space cache, allowing accurate shading and inpainting of newly disoccluded regions without hallucination. Our key contribution is a cache that is predictively filled with shading information for future views, amortized over multiple current frames. Novel views are synthesized by hierarchically traversing the cache front to back and refining stale or missing shading on the fly. Using a predictive, gradient-based scheduler that assigns  lifetimes for each tile, we enable adaptive shading updates under motion and dynamic lighting. \amulet decouples the rasterization and shading rate from the refresh rate of the display. In many scenarios, our cache can use a single shaded frame to synthesize multiple extrapolated frames with only a few localized updates. In a typical application, we extrapolate a 60 Hz shading rate to a 240 Hz display. \amulet achieves up to 250 Hz at 4K resolution and is competitive with state-of-the-art frame generation methods, including DLSS and neural-flow approaches, in multiple metrics. \amulet explores the design space of sparse layered image-space representation. It enables accurate, non-neural multi frame extrapolation with explicit handling of disocclusions. Our findings show that \amulet can extrapolate many more frames than contemporary methods with high quality, rivaling latency-bound frame interpolation methods with similar quality in many scenes.
\end{abstract}


\begin{teaserfigure}  
  \centering
  \includegraphics[width=\textwidth]{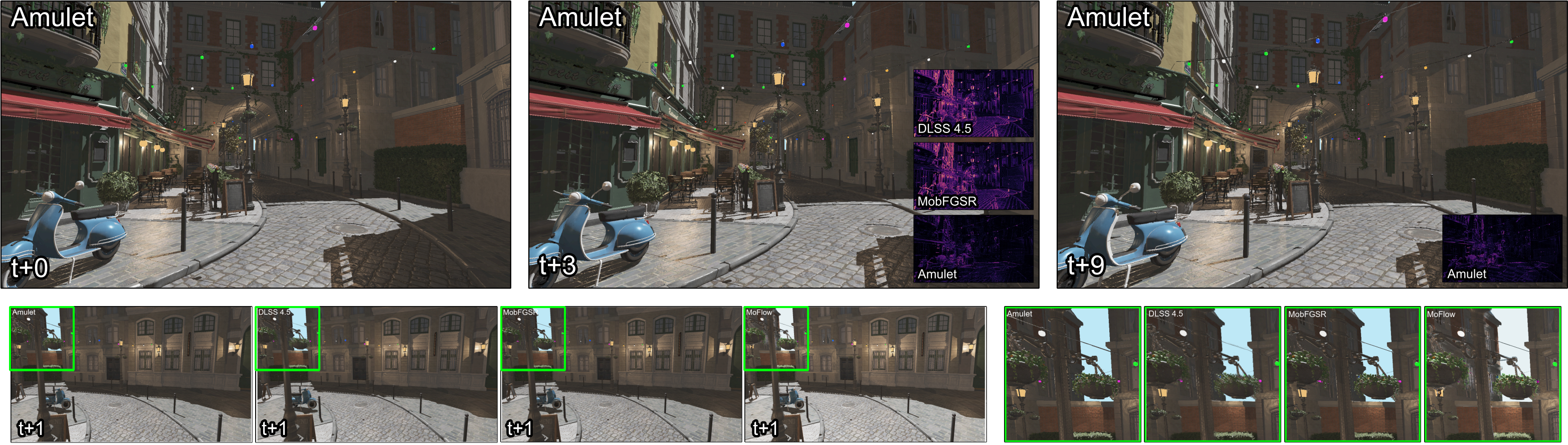}
  \caption{The top row shows three frames from the Bistro Exterior scene rendered with our frame extrapolation method: (left) The original keyframe, (middle) third extrapolated frame, (right) ninth extrapolated frame. The insets show FLIP error maps, with brighter colors indicating larger errors. The prediction quality is still high at nine extrapolated frames. The bottom row presents a side-by-side comparison of the first extrapolated frame from the Bistro Exterior scene produced by \amulet, DLSS 4.5, MobFGSR, and MoFlow. The frame extrapolation of our method, \amulet, outperforms neural and non-neural frame generation methods by filling disocclusions from its layered cache rather than hallucinating them. Competing methods are not able to produce nine extrapolated frames. The insets in the bottom row showcase ghosting and blurring artifacts present in all competing methods, whereas \amulet extrapolates frames without introducing visible artifacts. }
  \label{fig:teaser}
\end{teaserfigure}

\maketitle
\renewcommand{\shortauthors}{Künzel et al.}
\renewcommand{\shorttitle}{\mytitle}

\section{Introduction}

The demand for graphics throughput increases continuously. For example, current-generation game consoles operating at 4K/60\,Hz require $8\times$ more pixels than the previous generation operating at Full HD/30\,Hz. GPU improvements face cost and power constraints and cannot trivially keep up with this increase in demand. Exploiting spatiotemporal coherence for accelerated frame generation is the most promising mitigation strategy.

Current frame generation methods, such as Nvidia DLSS~\cite{dlss4}, MoFlow~\cite{MoFlow} or AMD FSR~\cite{fsr}, rely on neural networks for shading changes and relighting. Non-neural approaches, such as motion smoothing in SteamVR~\cite{Lang2018}, or MobFGSR~\cite{yang2024mobfgsr}, typically rely on reverse reprojection of the previous frame buffer. While efficient, these techniques operate on a single flat image and can handle larger disocclusions only imperfectly, often resulting in blurring, ghosting, or holes. A conventional frame buffer simply does not store enough information about the scene geometry to support extrapolation with novel disocclusions. 

To achieve frame extrapolation without disocclusion errors, the \textit{Talisman}~\cite{torborg1996} architecture proposed a layered rendering cache already three decades ago. Talisman replaced the linear (i.e., flat and dense) framebuffer with layered tiles and could handle significant disocclusions by adjusting and compositing its layers. Talisman hardware was never built, and the use of linear buffers to cache between graphics pipeline invocations still dominates today. 

In this paper, we introduce \amulet, a rendering pipeline inspired by Talisman, but implemented using contemporary GPU features. Our framework generalizes the concept of the disocclusion buffer introduced by~\citet{DisocclusionBuffer} into a general-purpose cache for efficient shading re-use. \amulet replaces the traditional elements of a deferred rendering system with alternatives of similar efficiency but enhanced flexibility. The conventional G-buffer is replaced with a shading cache composed of sparse tiles. The tiles are organized by object groups and depth layers. Refreshes of the geometry and shading samples contained in the cache are applied selectively and amortized over a variable number of frames. New frames are extrapolated by traversing the layers and compositing samples from non-empty tiles. Disocclusions are filled when tiles in subsequent layers are visited during traversal.

In particular, 
\amulet targets limitations that conventional history and motion vector based warping cannot resolve: These include novel disocclusions, such as peeking around corners or behind moving geometry, which cannot be reconstructed reliably from warped history and shading-only changes, such as moving shadow boundaries, which are difficult for motion-vector-based methods because static surfaces provide no motion vectors.
Unlike neural approaches that often require retraining or scene-specific tuning, \amulet can work with any scene directly.

In summary, we make the following contributions:
\begin{itemize}[leftmargin=9pt, topsep=0pt]
\item 
We introduce a sparse cache that replaces a linear framebuffer and anticipates disocclusions through its layered structure. 
\item
We describe a strategy for partitioning the cache so that parts of the stored geometry can be dynamically updated.
\item 
We present an adaptive shading method that amortizes shading over multiple frames and uses shading gradients to determine which parts of the scene must be updated first to anticipate rapid shading changes. 
\end{itemize} 

\section{Related work}
\label{chap:relatedWork}

Current work on frame extrapolation mostly uses G-buffers or other 2D caches (Section~\ref{sec:stup}). However, \amulet is more closely inspired by various 3D cache structures, which are better at filling in disocclusions with cached samples. These 3D caches can be categorized as view-independent (Section~\ref{sec:3dvicache}) or view-dependent (Section~\ref{sec:3dvdcache}).

\subsection{Shading caches in 2D}
\label{sec:stup}

Spatiotemporal upsampling can help reduce shading load by exploiting spatial and temporal coherence~\cite{Herzog2010}. If we want to synthesize entirely novel views, the classic approach is 3D warping~\cite{mark1997postrendering} of a conventional framebuffer or G-buffer. Reprojecting via a depth or motion vector buffer inevitably leads to undesirable disocclusions, limiting the extent of possible extrapolation. Even with sophisticated strategies for warping~\cite{bowles2012} or adaptive refresh~\cite{Sitthi2008}, disocclusions are hard to hide for longer reprojection distances.

One way to mitigate the disocclusion problem is by \textit{frame interpolation}, where additional upsampled frames are inserted to stabilize framerates between a previous and a current (or predicted) frame, typically using motion vectors~\cite{BidirSceneReprojection}. For example, information from the previous and next frames can be combined to generate intermediate frames with fewer artefacts~\cite{BidirSceneReprojection}; however, errors remain in highlights, shadows, reflections, transparencies, and thin objects.

Better image quality can be obtained by predicting or refining motion vectors with neural networks~\cite{briedis2023kernelbased}. Frame interpolation methods based on neural networks are now available through solutions such as DLSS~\cite{DLSS} and FSR~\cite{fsr}, although the technical details of these commercial solutions have not been revealed. Despite its popularity, frame interpolation has the fundamental drawback of adding additional latency, making it unsuitable for longer predictions or streaming applications that require keeping strict temporal bounds.

In contrast, \textit{frame extrapolation} methods generate novel views purely from past frames. Extrapolation thereby avoids additional latency, but it must handle disocclusions without the ability to peek into future frames. MobFGSR~\cite{yang2024mobfgsr} and its successor MobSS~\cite{Yang2025} compensates for this lack of disocclusion information using motion vectors and additional G-buffer channels. ExtraNet~\cite{guo2021extranet} and ExtraSS~\cite{wu2023extrass} train neural networks to infer missing information. 

Several recent methods improve this extrapolation strategy with additional measures. GFFE~\cite{Wu2024} opportunistically collects samples from multiple past frames to further increase the chance of being able to fill in disocclusions. MoFlow~\cite{MoFlow} addresses the restrictions associated with motion vectors to optimize the reuse of shading information for changing geometries, illumination, and translucent objects. StereoFG~\cite{Zuo2025} specializes in exploiting the cross-view coherence of stereoscopic image pairs. Its extrapolation considers coherence within frames, between stereo image pairs, and between successive temporal frames. EFXNet~\cite{Wu2025} explicitly predicts the magnitude of the extrapolation error and uses it to guide incremental shading toward areas in need of updates. 

However, the fundamental property of frame extrapolation is that missing information caused by disocclusions cannot be fully avoided. Our method therefore focuses on the timely generation of the missing shading information.

\subsection{View-independent shading caches in 3D}
\label{sec:3dvicache}

Only a cache layout capable of storing samples at arbitrary locations in 3D space can ensure that all relevant samples are present in the cache. Such alternative cache layouts can be roughly categorized into view-independent and view-dependent organizations. View-independent methods use either a hash table filled with individual shading samples~\cite{Ragan2011, liktor2012decoupled, Hladky2019b} or some form of texture space. 

The texture-space methods can be categorized into three groups: The scene is either pre-charted as a whole~\cite{Hillesland2016}, split into pre-charted primitive groups that are dynamically allocated at runtime~\cite{Mueller2018, baker2022}, or individual primitives are packed into a stratified atlas on the fly~\cite{Clarberg2014, Hladky2021}. Such texture-space organizations have been demonstrated to benefit applications such as radiance caching~\cite{Tole2002, Tatzgern2024} or streaming~\cite{Mueller2018, Hladky2021}. Obviously, these caches benefit from prioritized updates of areas with rapidly changing shading information~\cite{Mueller2021}. Less obvious is that frame extrapolation also benefits from the lower distortion of storing view-dependent sample patterns when reprojecting from a nearby view with a similar perspective~\cite{Hladky2021, Neff2022, FastAtlas}. 

This observation suggests advantages in using a view-dependent cache organization, such as the one used in \amulet. We next describe related work on view-dependent caches.

\subsection{View-dependent shading caches in 3D}
\label{sec:3dvdcache}

Since view-dependent sampling patterns benefit reprojection, a final category of related work organizes 3D shading caches directly using view-dependent layouts. Early techniques relied on rendering individual objects to \textit{impostor} buffers~\cite{Schaufler1996}, or proposed GPU hardware extensions~\cite{Diepstraten2004} or entirely novel hardware architectures like Talisman~\cite{torborg1996}. Other methods to address the third dimension (depth) use per-pixel linked lists~\cite{shade1998LayerDepthImages, Hladky2019, Franke2018}, depth peeling~\cite{Kravec2023, Kim2023, Franke2018}, or $k$-buffering~\cite{Bavoil2007}. More recent work generalizes impostors by including material information for on-the-fly reshading~\cite{ReshadableImpostors}, streaming-friendly representations~\cite{Hladky2022, Lu2025}, or supporting stereoscopic parallax in XR~\cite{ImpostorsXR}. Even some neural extrapolation methods rely on a peeled-layer-like structure as input for frame extrapolation~\cite{Wu2024}. 

Some memory layouts for caches rely on regular rather than scene-dependent depth spacing. Such a regular spacing can benefit shading caching~\cite{Schaufler1998, Mildenhall2019}, but also visibility computation~\cite{DisocclusionBuffer} or ray-tracing~\cite{Zeng2023}. The cache layout of \amulet also falls into this category. However, its tiled organization better exploits spatial coherence, surpassing previous approaches in efficiency. Adaptive shading and on-the-fly updating is a well known concept for framebuffers \cite{bishop94} and lends itself to shading caches beyond framebuffers.

\begin{figure}
    \centering
    \includegraphics[width=\columnwidth]{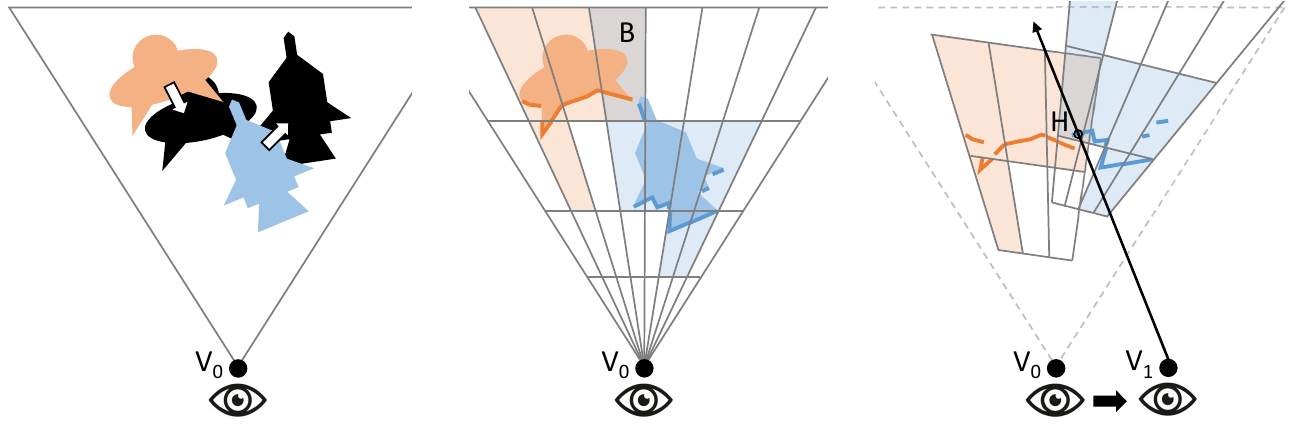}
    \caption{\amulet combines a cache generation stage with a frame extrapolation stage. (left) Two spaceships move toward the position shown in black, as seen from camera position $V_0$. (middle) Each spaceship is sampled into its own grid cell as indicated by the orange and blue contours. Only grid cells containing at least one sample of an object are stored. These grid cells are highlighted in light blue and orange. The grid cell labeled $B$ contains a blue and an orange part, which are separately included in their respective caches. (right) A new view is extrapolated from viewpoint $V_1$ after both the objects and the camera have moved. A ray from $V_1$ marches through the caches and intersects the non-empty cells on the way. The point $H$ of the blue spaceship is determined as the first hit.}
    \label{fig:extrapolate}
\end{figure}

\section{Method overview}
\label{chap:method}

Frame extrapolation based on reprojection, i.e., ``post-rendering warping''~\cite{mark1997postrendering}, is straightforward in principle, but complicated if samples are disoccluded or if the shading of visible samples changes. We address disocclusions by storing multiple layers of samples, and we handle shading changes by re-shading regions having high shading gradients in the temporal domain. Both techniques can be implemented by modifying a deferred rendering pipeline, which makes it easy to integrate frame extrapolation into existing rendering systems.

A deferred rendering pipeline typically consists of three rendering passes: First, a \textit{geometry pass} samples the geometric primitives into a G-buffer. For minimal bandwidth usage, a popular choice is to use a visibility buffer~\cite{Burns2013}, a lightweight G-buffer variant that contains only attachments for depth and primitive ID. Second, a \textit{shading pass} computes color information for the geometry samples. Third, a \textit{compositing pass} generates the final view by applying various post-processing steps, such as anti-aliasing or tone mapping. All three passes operate on a dense and flat buffer (i.e., a 2D texture) that stores exactly one sample per pixel location.

\amulet replaces all three passes with more flexible alternatives operating on a sparse cache (\autoref{fig:extrapolate}): Samples in \amulet are arranged in tiles covering $m\times m$ sample locations. Tiles, in turn, are arranged in $k$ layers covering consecutive depth intervals in the viewing frustum. Hence, the view frustum is subdivided into \emph{froxels} (frustum voxels)~\cite{evans_learning_2015}, and each froxel is associated with exactly one tile. Unlike a dense G-buffer, the \amulet cache is sparse: Only non-empty tiles are created and stored in the cache.

The geometry pass of \amulet samples the geometric primitives and writes visibility samples (depth and ID) to the cache. Initially, the cache is empty; tiles are created on the fly if the first sample is written. Each tile---and, therefore, each layer---has its own depth buffer. Consequently, the cache stores up to $k$ samples per pixel location. In other words, the cache resembles a variant of a $k$-buffer~\cite{Bavoil2007} storing at most one sample per layer. From two samples in the same layer that map to the same location in image space, the depth test retains only the closer one. If a primitive spans more than one layer, each of its samples will be assigned to exactly one tile in the layer corresponding to the sample's depth value.

The shading pass of \amulet iterates over the tiles that need shading and computes color values for all samples in the tile.  An important advantage of the \amulet shading pass is that shading reuse is maximized by choosing tiles with an adaptive refresh strategy.

The compositing pass of \amulet performs frame extrapolation. It generates a linear color buffer for a novel camera position. For every pixel location of the extrapolated frame, a ray is intersected with the froxel grid. If a froxel is non-empty, the associated tile is retrieved, and the ray is intersected with the depth buffer of the tile. If an opaque sample in the tile is hit, the ray terminates. Otherwise, the ray continues in the next layer.

We describe the basic rendering pipeline of \amulet in Section~\ref{sec:implementation}. Section~\ref{sec:dynamic} introduces the extension to multiple caches, which is the foundation for dynamic scenes. The strategy for adaptive shading is described in Section~\ref{sec:adaptive}, and the extension to scenes with transparencies is described in Section~\ref{sec:transparencies}.

\section{Rendering pipeline}
\label{sec:implementation}

We index in the froxel grid using the quantized coordinates of a sample point $\mathbf{p}=[x,y,z]^\top$, where $x, y$ are in screen coordinates, and $z$ is the linear depth between the near clip-plane $z_n$ and the far clip-plane $z_f$, i.e., $z_n<z<z_f$. The corresponding tile coordinate is
\begin{equation} \nonumber
C(\mathbf{p})=[c_x,c_y,c_z]^\top=\left[\left\lfloor \frac{x}{m}\right\rfloor, \left\lfloor \frac{y}{m}\right\rfloor, \left\lfloor k\frac{\log(z-z_n+1)}{ \log(z_f-z_n+1)}\right\rfloor\right]^\top
\label{eq:one}
\end{equation}
for a cache with $k$ layers and logarithmic spacing (i.e., the layers become exponentially wider). 

The cache is implemented as a sparse structure~\cite{Kraus2002} in GPU memory. Tiles containing at least one sample are placed in a storage buffer that allows reads and writes with random access. To access the sparse tile storage, we use a software-implemented page table with one entry for every possible froxel grid position. An entry in the page table points to the tile address if the tile exists; otherwise, it is null. When the first sample is written to a tile, the tile is atomically allocated in the storage buffer, its depth set to $z$, and its page table entry pointed to the tile address. 

\subsection{Geometry pass}

The geometry pass fills the cache with visibility samples of all primitives in the scene, up to $k$ samples distributed across $k$ layers for a given pixel location. Each visibility sample is a 64-bit value that encodes depth in the higher 32 bits and primitive ID in the lower 32 bits. Depth is not expressed globally, but is layer-relative, to increase precision and reduce depth-fighting. Layer-relative means that a depth value is transformed into a range of [0, 1] covering the width of the layer in which the sample is stored. The layer-relative depth of a sample in layer $i$ is $(z-z_i)/(z_{i+1}-z_i)$, where $z$ is the global depth of the sample and $z_i$ the depth of the layer.

The geometry pass feeds the scene primitives to the hardware rasterizer to generate the visibility samples. However, the fragment shader does not use conventional raster operations to access the framebuffer. Instead, it implements a depth buffer operation in the tile storage with an atomic minimum operation~\cite{Karis2021, DisocclusionBuffer}.

\subsection{Shading pass}
\label{sec:methodtiledshading}

Tile shading is implemented as a compute shader dedicating one thread per sample. Similarly to the use of a conventional visibility buffer~\cite{Burns2013, Karis2021}, the shading reconstructs the world-space position per sample from the pixel location and depth value and uses the primitive ID to determine barycentric coordinates and interpolated attributes. We apply a physically based rendering model with normal mapping and ray-traced shadows and path-traced global illumination in selected scenes. 

\subsection{Compositing pass}
\label{sec:methodfe}

Unlike most reverse reprojection methods, which suffer from the ambiguities inherent to motion vectors, our frame extrapolation approach operates directly on the geometric structure of the cache. For every pixel location of a novel view, a view ray is cast through the cache, and all layers are traversed in front-to-back order. The ray origin and direction are computed by transforming the current extrinsic camera parameters into the view space that was used to create the cache. 

For each layer, we determine the intersection points of the ray with the enclosing image planes and map these points to tile coordinates. A 2D DDA enumerates the tiles and sample locations along the ray segment within the layer. Empty tiles, which are indicated by a null pointer in the page table, can be skipped. For non-empty tiles, we compare the depth values stored in the visibility samples to the ray depth. A hit is found if the view ray depth is larger than the one of the sampled depth values from the cache. 

To accelerate the traversal of the froxel grid, we employ empty space skipping informed by a hierarchical occupancy mask. A bit set in the mask indicates a grid region that contains at least one non-empty tile. The occupancy mask structure has two levels:

\textbf{Level-1} masks aggregate occupancy information for a small block of froxels.

\textbf{Level-2} masks aggregate occupancy information for a larger block of froxels, corresponding to multiple level-1 masks.

Level-1 masks are intended for fine-grained skipping in regions that contain geometry samples. The level-1 region size is intentionally chosen to be small so that multiple froxels can be skipped during the traversal of a layer. A larger region size decreases the potential for skipping, because the probability of occupancy increases with the size of the region. The masks are generated with a compute shader using one thread per froxel that sets its corresponding occupancy bit in the mask using an atomic OR-operation.

During frame extrapolation, we use the occupancy mask for skipping as follows: For each coordinate generated by the DDA, the procedure first tests the layer-1 mask to determine if an entire region of froxels can be skipped. If not, it inspects the page table entries to determine if a tile exists. If it does, the procedure marches along the ray through the tile's depth buffer to find a possible intersection.

Furthermore, a group of layers contained in the mask can be skipped in advance by checking all cells the ray enters for occupancy.

Level-2 masks cover regions of much larger granularity and allow skipping large portions of layers at once in sparsely occupied areas. The size of a level-2 mask is chosen to fit exactly in a single cache line to optimize the utilization of memory bandwidth. The level-2 masks are generated using a similar shader as for level-1 masks, but inspecting groups of level-1 masks instead of froxels.

With the level-2 masks, we add another entry at the top of the space skipping hierarchy. To keep this test concise, we do not set up a 3D DDA for intersecting the ray with the exact regions covered by level-2 masks. Instead, we use the coarser, but faster method of computing the bounding box of the regions intersected by the ray and test it against the level-2 mask. This method takes into account the fact that the start and end points of a ray are often contained in the same or neighboring level-2 regions, making it faster overall.

\begin{figure}
    \centering
    \includegraphics[width=\columnwidth]{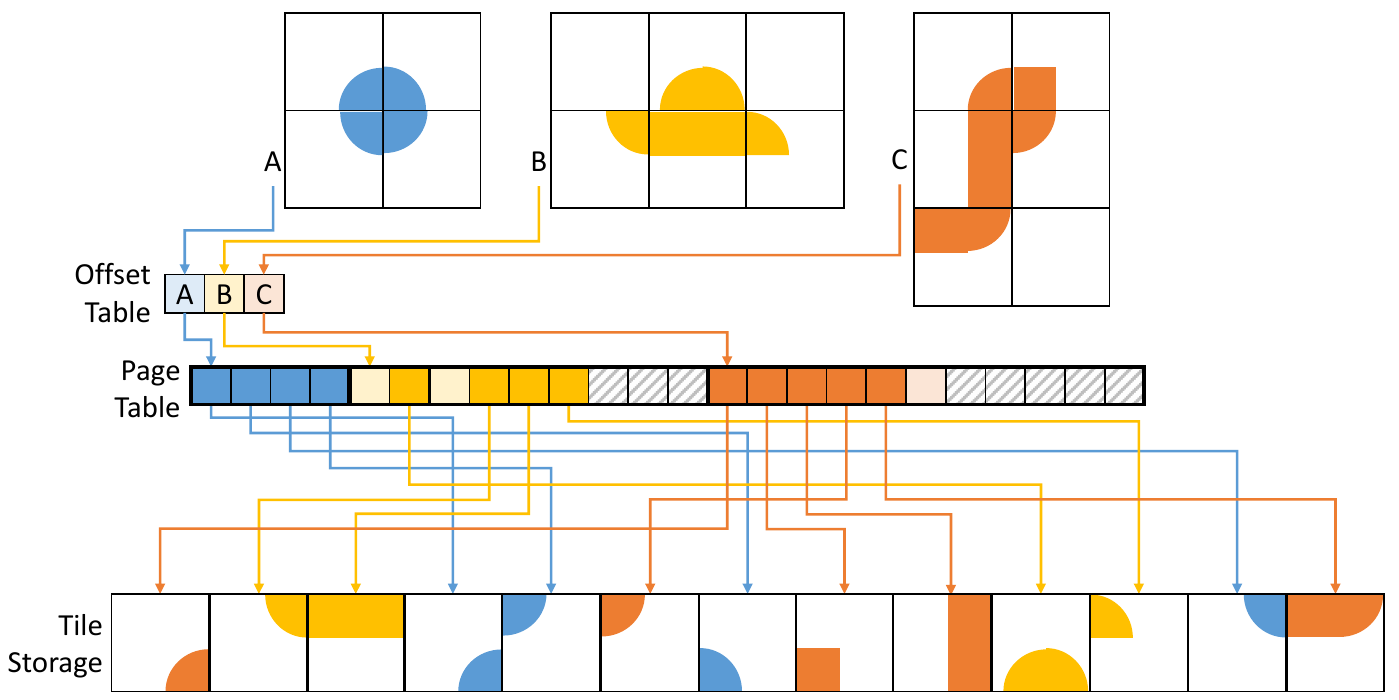}
    \caption{A scene with three objects, A, B, and C. Each object is stored in its own cache. To access its cache, an object refers to its entry in the offset table to determine where its page table entries start. Each page table entry points to an element in the tile storage corresponding to a non-empty froxel of the space occupied by the object.}
    \label{fig:mcache}
\end{figure}

\section{Dynamic scenes}
\label{sec:dynamic}

In dynamic scenes, separate caches let us refresh each object independently, thereby maximizing cache reuse. The first object cache is dedicated to the static part of a scene, and additional caches can be used for dynamic objects or object groups. For that purpose, we split the lower 32~bits of a visibility sample into a subcache ID (5~bits) and a primitive ID (27~bits). This encoding supports 32 separate caches with 134\,M primitives each, sufficient for production scenes.

The cache for the static scene derives the dimension of its froxel grid from an enlarged camera frustum that encloses the relevant part of the scene visible from future camera positions~\cite{Voglreiter2023}. In contrast, caches for dynamic objects employ a tighter froxel grid that only encloses the contained object. Such a froxel grid is constructed from an axis-aligned bounding box of the object in the view space given at the time of cache generation. 

We extend the page table structure to provide room for multiple object caches (\autoref{fig:mcache}). An additional table stores the offset into the page table for each object, leading to a three-level hierarchy (offset table, page table, tile storage). Page table reallocations can become necessary if the cache size of the object changes during animation and the allocated portion of the page table overflows when trying to refresh an object cache. Therefore, the size of each object's page table portion is taken from the actual size of the object cache and padded to the next power of two. Padding ensures that page table reallocations are rare.

In the compositing pass, each view ray of the extrapolated frame is transformed into the view space of each object cache. View ray warping requires tracking of the current and initial model matrices during cache generation of the object contained in the subcache. The warp matrix is $\mathbf{W}(t_n) = \mathbf{V}(t_0) \cdot \mathbf{M}(t_n) \cdot \mathbf{M}(t_0)^{-1} \cdot \mathbf{V}(t_0)^{-1}$, where $\mathbf{V}$ is the view matrix, $\mathbf{M}$ is the model matrix, $t_0$ is the time of cache generation, and $t_n$ is $n$ frames later. The origin and direction of the view ray are warped using $\mathbf{W}(t_n)^{-1}$. As cache depths are not valid under object transformations, a reconstructed view space position is warped using $\mathbf{W}(t_n)$ to get a depth value for the transformed object. Otherwise, the depth sorting of samples would be incorrect.

Among the first hits in all caches, the sample with the smallest ray distance is kept. The caches for moving objects are small, and the potential for empty space skipping is negligible; therefore, we do not generate occupancy masks for moving objects and skip only tiles marked as empty in the page table.

\section{Adaptive shading and visibility}
\label{sec:adaptive}

\amulet is designed to amortize shading costs over multiple frames. The samples in the \amulet caches represent both the geometry and the shading of the scene. \textit{Shading updates} are handled at the granularity of individual tiles. A shading update for a tile is required when the shading of at least one sample of the tile has changed too much. We assume that computation is dominated by shading invocations and, therefore, aim to reuse shading efficiently.

\textit{Geometry updates} must be handled at the coarser granularity of individual object caches. Updating the geometry of an object cache implies rebuilding it entirely. Reprojecting the old shading into the new cache is not possible, as it would degrade the shading quality too much. Therefore, we discard the shading information when rebuilding a cache and shade it anew before the first use. 

\begin{figure}
    \centering
    \includegraphics[width=0.99\columnwidth]{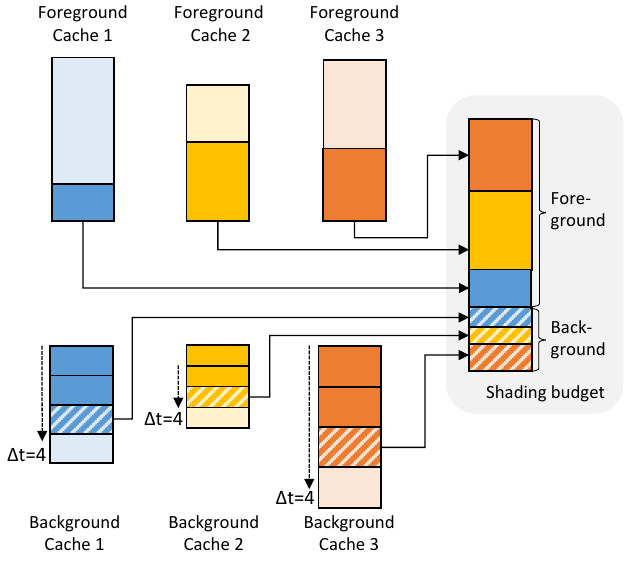}
    \caption{The selection of tiles for adaptive shading in a scene decomposed into three caches. Each consists of a foreground and a background cache. The foreground caches are updated on demand and immediately used for shading. Tiles are shaded if they are currently visible and their time to live has expired. The background caches are prepared ahead of time and shaded over multiple frames $\Delta t$ with a round-robin strategy. Cumulatively, the  tiles selected for shading must fit into a given shading budget.}
    \label{fig:adaptive}
\end{figure}

Shading one or multiple object caches at once leads to a spike in shading load. We address this problem by double-buffering each cache. The foreground cache contains the shading used for frame extrapolation. The background cache is created ahead of time using a predicted scene state, including camera position and object transforms. The shading of the foreground cache and the generation of the background cache are continuously updated, but with different strategies (\autoref{fig:adaptive}).

\subsection{Updating the foreground cache} 

Shading updates are applied to samples with rapidly changing shading to ensure that a foreground cache can be used for extended periods of time. \citet{Mueller2021} show that shading gradients can be robustly estimated from finite differences between shader invocations and that a substantial portion of the samples can tolerate extended periods of delayed reshading without noticeable artifacts. 

Each tile in a foreground cache is assigned a \textit{time to live} (TTL) based on its shading gradient. Initial shading gradients are determined by shading tiles twice, once regularly and a second time at a 16$\times$ reduced resolution with the predicted future scene state. When a tile is reshaded, its new shading gradient is determined on the fly by comparing the new shading values to the old ones before overwriting the old shading values. The magnitude of the shading gradient is determined as the maximum over all samples in the tile. A convolution spreads the maximum magnitude over neighboring tiles, creating a safety region of conservative estimates. A threshold on the gradient magnitude is then used to flag a tile for reshading by zeroing the TTL above the threshold. 

We compute the list of tile candidates for reshading by running a modified compositing pass \textit{before} the shading pass. The modified pass traverses the froxel grid to identify the visible tiles that map to pixels of the extrapolated frame. Tiles will not be considered for reshading until they become visible. A tile is reshaded only if it is visible and its TTL has expired. Foreground updates are given the highest priority. In our evaluation, the load caused by foreground updates was consistently light compared to background updates and usually almost constant for typical movement. 

\subsection{Updating the background cache}

Meanwhile, a background cache is created with refreshed geometry corresponding to a predicted scene state for a future frame when the caches will be swapped. We considered deriving the number of frames until the background and foreground caches should be swapped from the speed of object motion. However, if object speeds are moderate and most objects remain visible, a fixed amortization period $\Delta t$ proved to be sufficient.

Our predictive scheduler amortizes the shading load over the generation time of the cache, yielding a roughly constant per-frame load for background cache generation. The background cache only shades tiles that are visible given the predicted scene state. Prediction errors and camera or object motion during extrapolation from the cache are handled by adaptive foreground cache updates.

\subsection{Semitransparent scenes}
\label{sec:transparencies}

\begin{figure}[t]
    \centering
    \includegraphics[width=0.99\columnwidth]{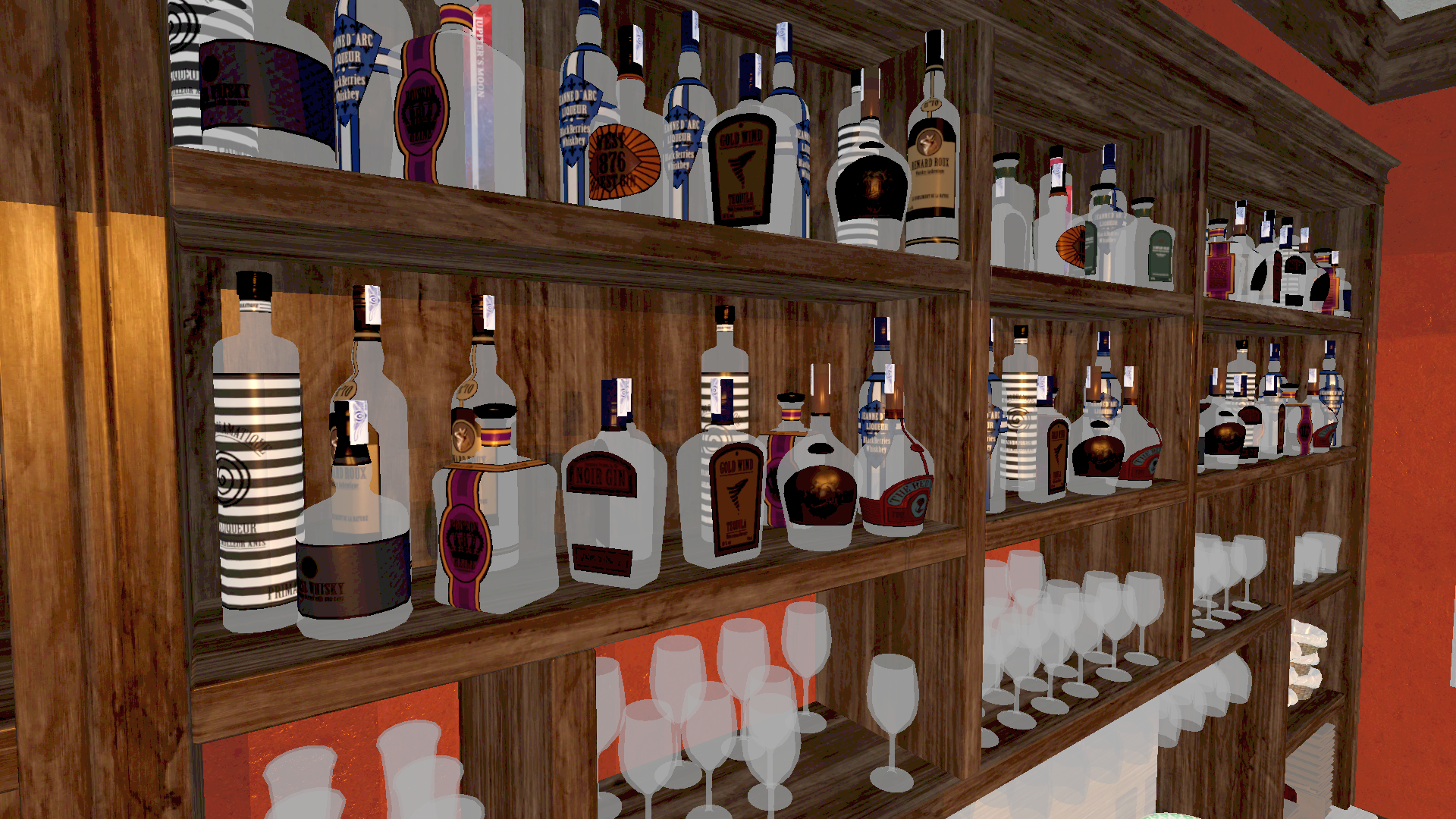}
    \caption{With our method, the glassware in the Bistro Interior scene can be extrapolated without destroying the semitransparent appearance. Objects seen through glassware change depending on the observer's perspective.}
    \label{fig:glass}
\end{figure}

To compactly store a variable number of semitransparent samples per tile, we extend the page table with an additional hierarchical level that provides the software equivalent of page table virtualization. This modification ensures that the memory footprint of the page tables remains reasonable even if semitransparent samples accumulate densely along a view ray. For semitransparent objects, the page table entry for a non-empty tile points to a \textit{tile record} that contains $P$ pointers, each of which may point to a transparency tile. Up to $P$ semitransparent samples can be stored for the same sample location in a given froxel. Semitransparent tiles are only created as needed; unused pointers in the tile record are set to null. 

\begin{figure*}[t]
    \centering
    \includegraphics[width=0.24\textwidth]{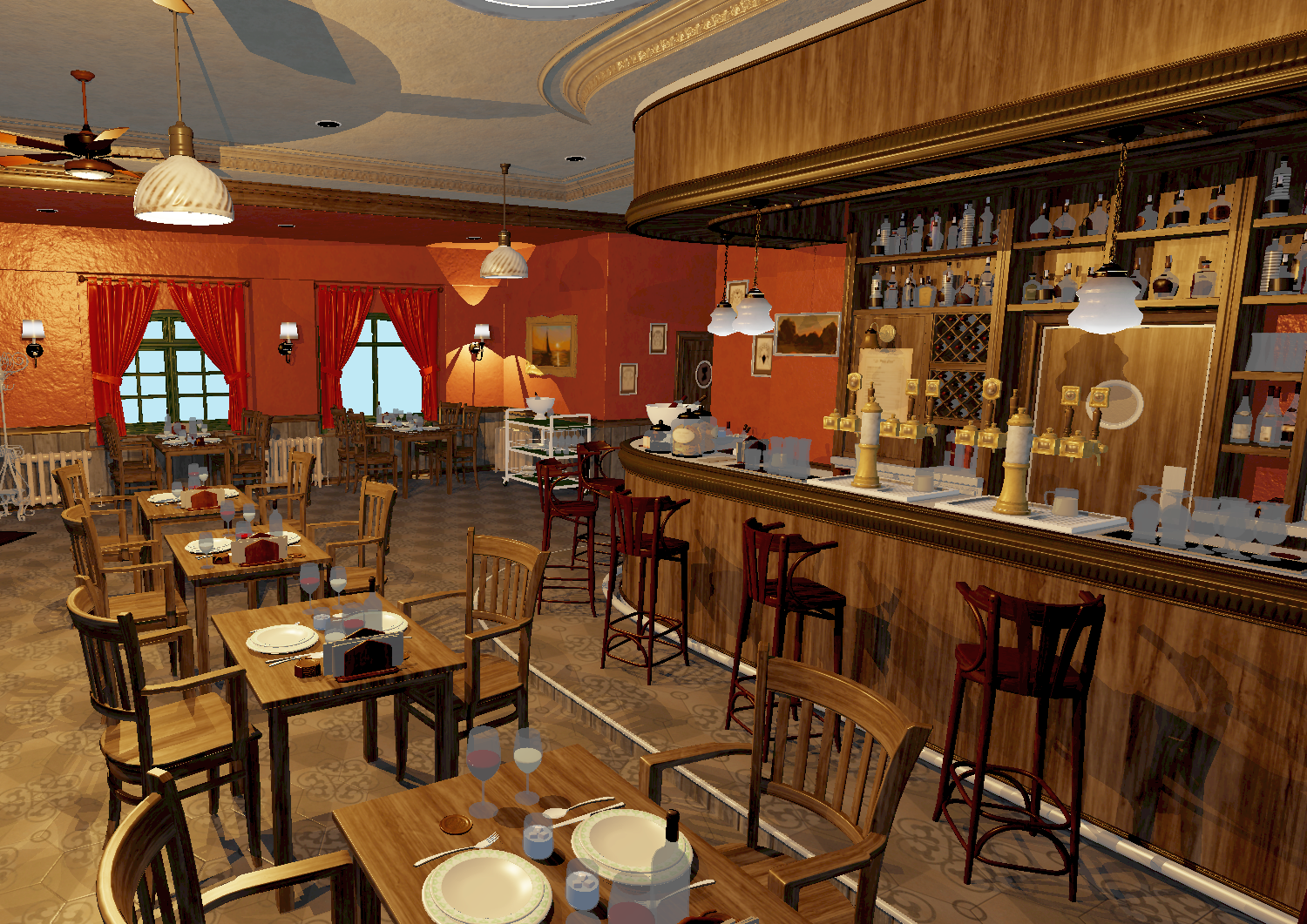}
    \hfill
    \includegraphics[width=0.24\textwidth]{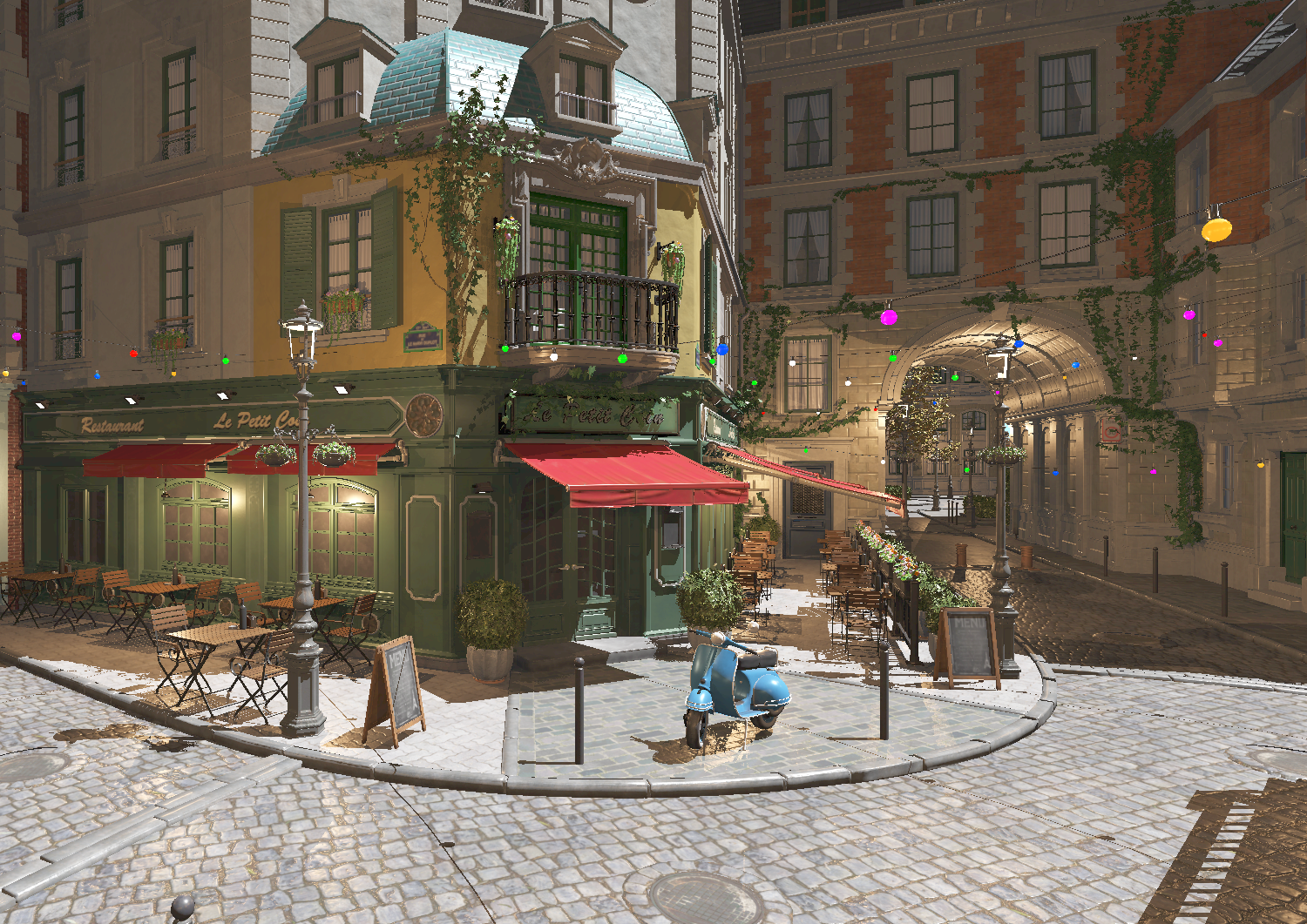}
    \hfill
    \includegraphics[width=0.24\textwidth]{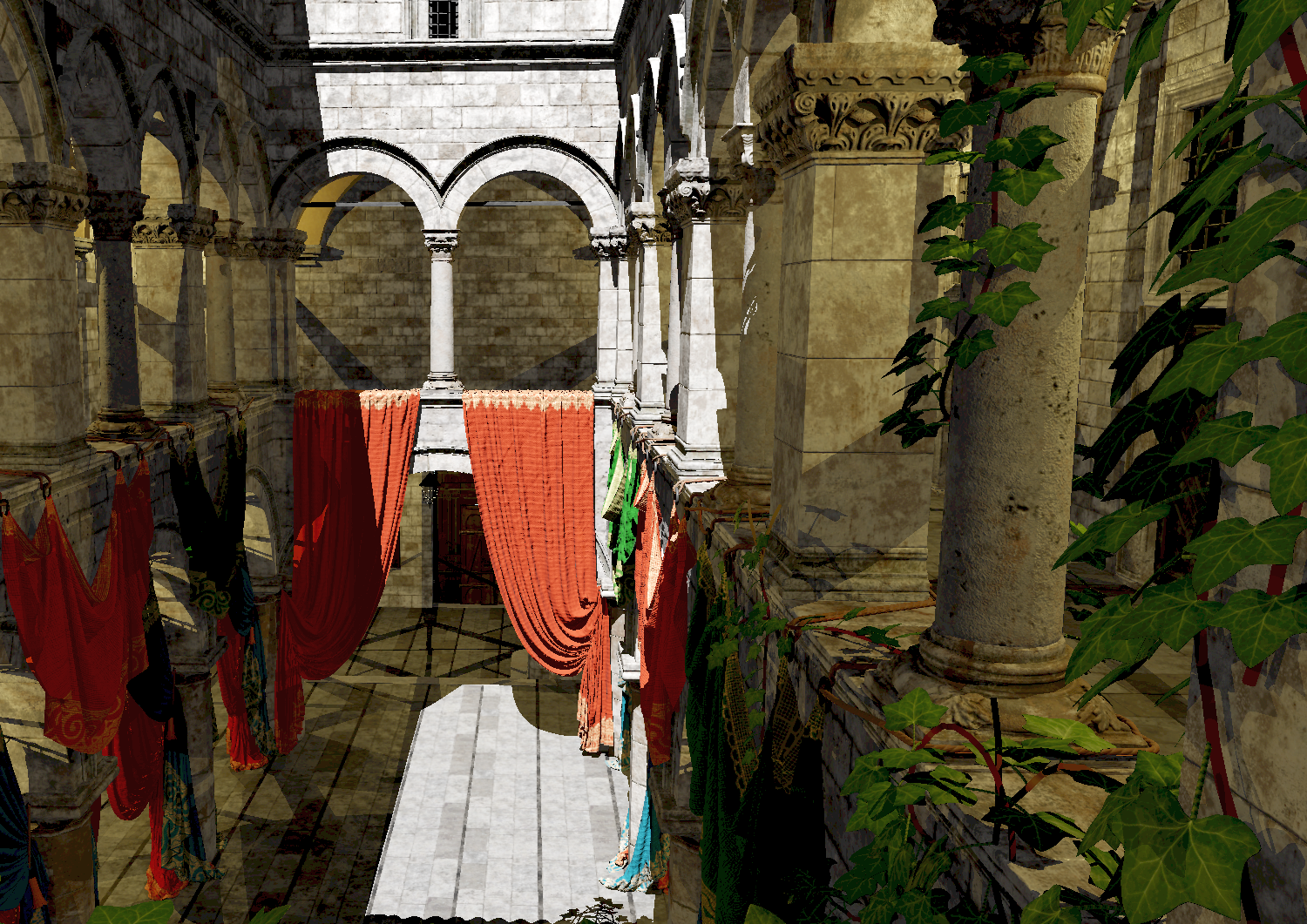}
    \hfill
    \includegraphics[width=0.24\textwidth]{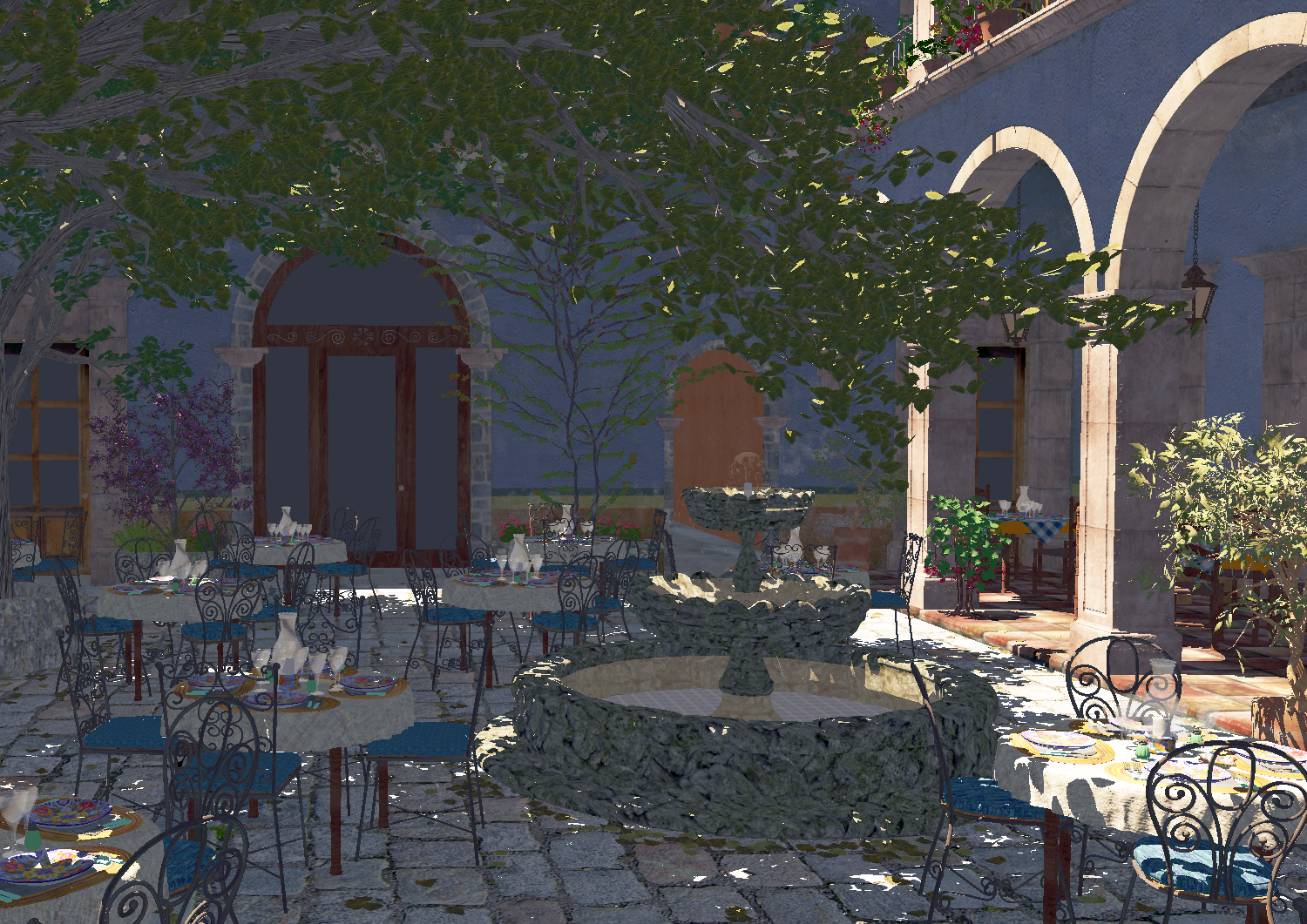}
    \caption{Our test scenes, from top left to bottom right: Bistro Interior, Bistro Exterior, Intel Sponza, and San Miguel.}
    \label{fig:scenes}
\end{figure*}

Writing to a cache for semitransparent samples is only marginally more complex than writing to a cache for opaque samples: When the first sample in a tile is encountered, but its page table entry is still null, the tile record is atomically created and referenced in the page table entry. Before writing a sample, the tile record is searched for non-empty pointers. The tiles referenced by non-empty pointers are searched sequentially until an empty sample (indicated by a depth value of one) is found or all existing tiles have been checked. If no empty sample is found, a new transparency tile is atomically allocated, and the tile record is updated. The sample is atomically written to the desired location. If $P$ samples have already been stored in a given location, any new samples for the same location are discarded. While efficient, this strategy requires choosing a suitable value of $P$ to match the expected overdraw of semitransparent fragments in advance. As shown in \autoref{fig:glass}, We did not observe any visual artifacts from discarding overflowing semitransparent samples in our test scenes . 

The semitransparent samples associated with a froxel are recorded in the order of appearance. We found that depth sorting these fragments before blending was too slow. Therefore, we adopted weighted blended order-independent transparency~\cite{WBOIT} to determine pixel colors during the compositing pass. Opaque and semitransparent objects are kept in two separate static caches. First, the compositing runs on the opaque cache to obtain an initial linear color and depth buffer. Second, the compositing traverses the semitransparent cache at most to the depth indicated in the opaque depth buffer. All semitransparent fragments in the currently visited layer are aggregated using WBOIT. If alpha becomes saturated, the ray is terminated early. Otherwise, traversal proceeds with the next layer. Consecutive layers are combined with conventional front-to-back blending instead of WBOIT. Finally, the semitransparent and opaque portions are blended.

\section{Results}
\label{sec:results}

\begin{figure*}
    \centering
    \includegraphics[width=0.99\linewidth]{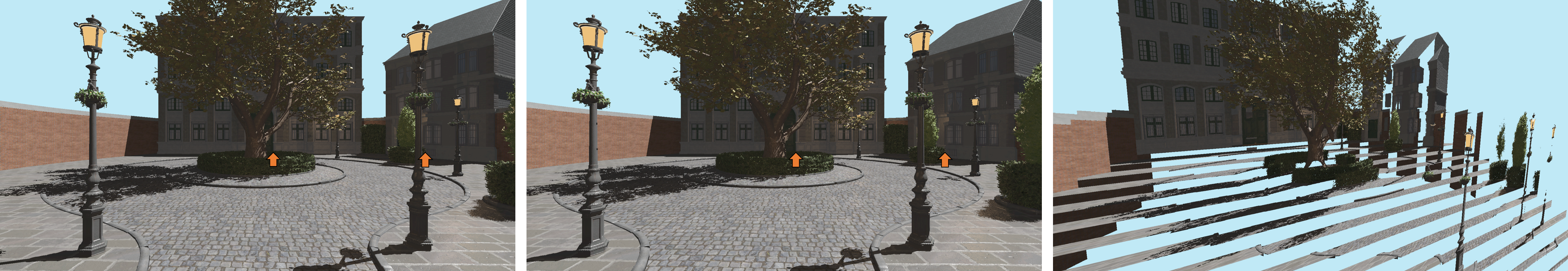}
    \caption{The \amulet cache can cover substantial camera movements with extrapolated images. The left and middle images show significant disocclusions on the buildings in the background behind the tree and the street lamp. These frames are extrapolated from the layered cache structure shown on the right.}
    \label{fig:extrapolation_example}
\end{figure*}

We implemented \amulet in the Falcor rendering framework~\cite{Falcor} using Vulkan bindings. Our cache implementation extends the disocclusion buffer introduced by~\citet{DisocclusionBuffer} with shading data for frame synthesis, multi-object warping and adaptive on-the-fly reshading. All benchmarks were recorded on a PC (CPU: AMD Ryzen 7 9700X, RAM: 64 GB, GPU: Nvidia GeForce RTX 5090, OS: Windows 11). We tested \amulet with two configurations: \textit{Full HD} (resolution 1920\,\ti\,1080) and \textit{4K} (resolution 3840\,\ti\,2160). Our implementation uses $N=64$ layers. This choice provides a good tradeoff between low memory footprint and depth quantization error. Tile sizes were chosen as 16\,\ti\,16 for Full HD and 32\,\ti\,32 for 4K. This choice is motivated by the fact that smaller tiles increase the page table size and make empty space skipping less effective. Conversely, larger tiles save memory for the page table hierarchy, but reduce the tile fill rate and waste tile storage. The shading cost remains independent of the chosen tile size. Occupancy masks for empty space skipping are sized as follows: Level-1 masks cover $4 \times 2 \times 2$ froxels. Level-2 masks cover $32\times16\times32$ froxels (i.e., $8 \times 8 \times 16$ level-1 masks). 
The occupancy mask covers $4 \times 2 \times 2$ froxels on Level-1 and $32\times16\times32$ froxels on Level-2.

We used four well-known test scenes: Bistro Interior, Bistro Exterior, Intel Sponza, and San Miguel (\autoref{fig:scenes}). An extrapolation example is shown in ~\autoref{fig:extrapolation_example}. Bistro Interior is equipped with animated chairs (\autoref{fig:big_comparison_1}). Large scenes with high depth complexity are more challenging for our system. All tests use a scene-dependent FOV. The extended resolution is 25\% larger than the original image resolution. Recordings were made using prerecorded camera paths, moving at 2-3 m/s, visiting all parts of the scene. Results were averaged over all frames.

\subsection{Memory consumption}

\begin{table}[t]
\centering
\caption{Memory consumption for the cache in MB averaged over all scenes for resolutions \textit{Full HD} and \textit{4K}. Memory consumption ranges from 85--700\,MB. Note that the number of tiles is comparable in both cases, because the tile size in \textit{4K} is doubled. }
\label{tab:memory_usage}
\begin{tabular}{lrr}
\toprule
\textbf{Scene} & \textbf{Full HD} & \textbf{4K} \\
\midrule
Bistro Interior & 87.53  & 285.01 \\
Bistro Exterior & 112.07 & 385.48 \\
Sponza          & 190.97 & 700.45 \\
San Miguel      & 145.02 & 518.58 \\
\bottomrule
\end{tabular}
\end{table}

The caches consume between 85--700\,MB, depending on the scene and the chosen resolution (\autoref{tab:memory_usage}). While this memory footprint is about 4\ti\ to 10\ti\ higher than that of a flat visibility buffer, overall memory requirements stay well below 1\,GB and are easily handled by any modern GPU. Memory consumption is highest when the camera is close to a view-filling object, such as a wall or a tree with dense foliage, making layers close to the camera densely occupied. These measurements correspond to about 22,000 to 58,000 tiles. Memory footprint and tile count depend on the resolution and layer count, while being rather insensitive to variations of the scene geometry. Therefore, the extrapolation effort is largely decoupled from scene complexity, which allows providing reliable bounds for render budgets. 

\begin{table}[t]
\centering
\begin{tabular}{llrrr}
\toprule
\textbf{Res.} & \textbf{Layers} & \textbf{Tiles} & \textbf{Mem. (MB)} & \textbf{Time (ms)} \\
\midrule
\multirow{3}{*}{Full HD}
& 32  & 24,847 &  83.82 & 3.54 \\
& 64  & 30,798 & 112.14 & 3.46 \\
& 128 & 39,143 & 158.38 & 3.56 \\
\midrule
\multirow{3}{*}{4K}
& 32  & 24,990 & 303.87 & 6.60 \\
& 64  & 31,049 & 385.78 & 6.60 \\
& 128 & 39,616 & 507.96 & 6.60 \\
\bottomrule
\end{tabular}
\caption{Cache generation times and memory consumption for different numbers of layers in the cache. All results are averaged over all caches produced during the camera sequence in the Bistro Exterior scene. Tests were conducted with 32, 64, and 128 layers. Cache generation does not depend on the number of layers. The average tile count and memory consumption increase sublinearly with the number of layers. Tile memory has the largest impact on the memory footprint. \amulet is able to fill the layered cache with moderate memory cost.  }
\label{tab:layer_ablation}
\end{table}

~\autoref{tab:layer_ablation} shows how the number of layers affects overall memory consumption. 
Due to the sparse design of our cache structure, the number of tiles in the cache—and therefore the required GPU memory—scale sublinearly with the number of layers. Empty regions are not represented in the cache.

The average fill rate of the page table is 1.3--3.6\%, which underlines the need for a sparse cache. Tiles are typically filled between 75--88\%, while being insensitive to resolution. 

\subsection{Performance}

\begin{table}[t]
\centering
\caption{Per-frame runtime comparison (in ms) between \amulet and a baseline deferred renderer. Although \amulet incurs an additional cost for composition, it achieves higher performance by amortizing the geometry and shading costs over multiple frames.}
\label{tab:runtime_comparison}
\begin{tabular}{llcrrrr}
\toprule
\textbf{Scene} & \textbf{Res.} & \textbf{Meth.} & \textbf{Geom.} & \textbf{Shad.} & \textbf{Comp.} & \textbf{Tot.} \\
\midrule
\multirow{4}{*}{\makecell[l]{Bistro\\Interior}}
& \multirow{2}{*}{Full HD} & \amulet   & \textbf{0.35} & \textbf{3.77}  & 1.00 & \textbf{5.12} \\
&                          & Deferred & 2.06 & 11.89 & \textbf{--} & 13.95 \\
& \multirow{2}{*}{4K}      & \amulet   & \textbf{0.73} & \textbf{3.15}  & 1.90 & \textbf{5.78} \\
&                          & Deferred & 2.24 & 11.26 & \textbf{--} & 13.50 \\
\midrule
\multirow{4}{*}{\makecell[l]{Bistro\\Exterior}}
& \multirow{2}{*}{Full HD} & \amulet   & \textbf{0.62} & \textbf{3.93}  & 0.79 & \textbf{5.34} \\
&                          & Deferred & 1.36 & 11.59 & \textbf{--} & 12.95 \\
& \multirow{2}{*}{4K}      & \amulet   & \textbf{1.38} & \textbf{5.50}  & 1.41 & \textbf{8.29} \\
&                          & Deferred & 1.70 & 22.84 & \textbf{--} & 24.53 \\
\midrule
\multirow{4}{*}{Sponza}
& \multirow{2}{*}{Full HD} & \amulet   & \textbf{1.28} & \textbf{3.10}  & 0.49 & \textbf{4.87} \\
&                          & Deferred & 3.82 & 11.46 & \textbf{--} & 15.28 \\
& \multirow{2}{*}{4K}      & \amulet   & \textbf{2.66} & \textbf{2.95}  & 1.01 & \textbf{6.61} \\
&                          & Deferred & 4.41 & 11.29 & \textbf{--} & 15.70 \\
\midrule
\multirow{4}{*}{\makecell[l]{San\\Miguel}}
& \multirow{2}{*}{Full HD} & \amulet   & \textbf{0.82} & \textbf{3.41}  & 0.84 & \textbf{5.07} \\
&                          & Deferred & 3.00 & 9.80  & \textbf{--} & 12.80 \\
& \multirow{2}{*}{4K}      & \amulet   & \textbf{1.57} & \textbf{3.07}  & 1.33 & \textbf{5.98} \\
&                          & Deferred & 3.20 & 8.42  & \textbf{--} & 11.62 \\
\bottomrule
\end{tabular}
\end{table}

Runtime was measured using GPU timers that directly wrap the interpolation/extrapolation steps during rendering. As can be expected, the time to generate a cache is proportional to the number of samples. ~\autoref{tab:runtime_comparison} shows the breakdown of the passes required for cache generation. In this test, we use a period of four frames for extrapolation in \amulet. We compare the performance to a deferred renderer that shades its G-buffer in every frame. 

The shading load is dominated by the number of rays in ray-tracing. The shading load (by increasing indirect light samples) was selected so that the deferred renderer could sustain slightly more than 60\,fps, except for Bistro Exterior in 4K, where it was closer to 30\,fps. With this shading load, \amulet outperforms deferred rendering overall by a factor of about 3--4\ti. 

The geometry pass of \amulet is about 2\ti\ as expensive as the G-buffer pass of the deferred renderer. Amortized over a period of four frames, the geometry pass of \amulet is already twice as fast as that of the deferred renderer. The number of layers in the cache does not influence the execution time of the geometry pass (\autoref{tab:layer_ablation}), because the computational load scales with the number of fragment shader invocations, which depends only on the depth complexity of the scene and not on the depth complexity of the layered cache. 

The compositing pass of \amulet, which performs both frame extrapolation and on-demand re-shading of the foreground caches, is always below 1\,ms for Full HD and below 2\,ms for 4K. In other words, frame extrapolation times increase sublinearly with resolution. Empty space skipping is essential for this outcome, because it provides a speedup of 3--4\ti. 

The shading pass of \amulet requires 2--6\,ms per frame, which roughly matches the shading load of the deferred renderer divided by the period of four frames, as intended.

\subsection{Subcache performance}

\begin{figure}[!t]
    \centering
    \includegraphics[width=\columnwidth]{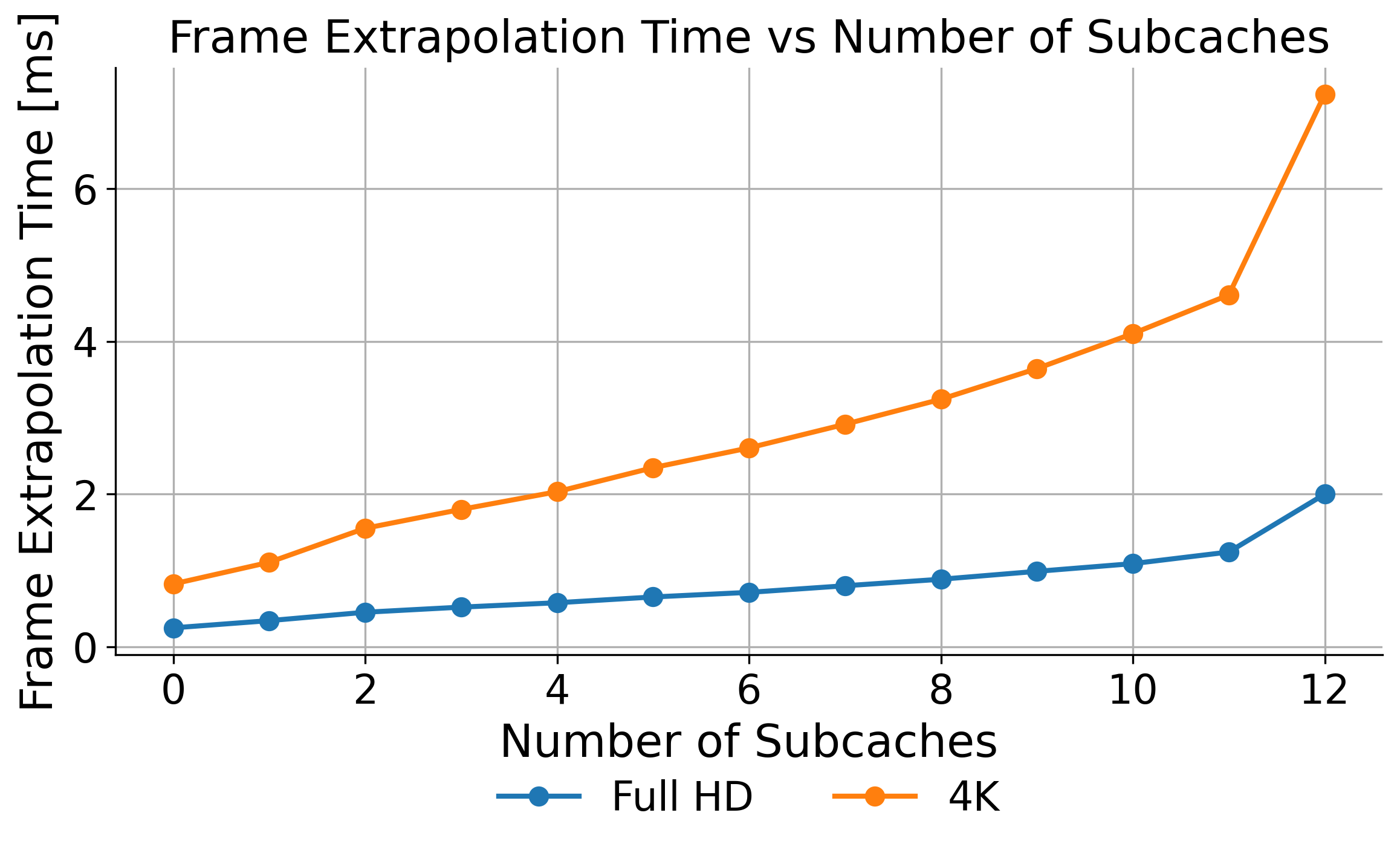}
    \caption{Frame extrapolation times (in ms) are shown for an increasing number of subcaches ranging from 0 to 12 inside Bistro Interior. Frame extrapolation times scale linearly with the number of subcaches.}
    \label{fig:subcache_ablation}
\end{figure}

We evaluate the performance implications of using multiple subcaches within a scene. We placed twelve rotating chairs in the Bistro Interior scene, each in a separate subcache. The number of layers in each subcache depends on its distance from the camera, but we ensure that there were at least 16 layers to prevent visual artifacts during reprojection. The extrapolation times are plotted in~\autoref{fig:subcache_ablation}. 
The results show that the extrapolation times increase linearly with the number of caches, because frame extrapolation must march sequentially through the caches. We tested parallelizing this process within a compute shader, but the performance was severely impacted by the scattered memory accesses between subcaches, as subcaches are stored one after another in the cache. Furthermore, subcaches do not benefit heavily from empty space skipping, because they are typically more densely filled than the cache containing the static scene. The number of subcaches must be selected with the target budget per frame in mind.

\subsection{Comparison to state of the art}

We compare the rendering times and the visual quality of extrapolated frames using ground-truth images created with standard deferred rendering for every frame. ~\autoref{tab:image_quality_comparison} shows visual errors per scene measured by several image metrics: PSNR, SSIM~\cite{SSIM}, FLIP~\cite{Flip}, and LPIPS~\cite{LPIPS}.

\paragraph*{Baselines}
We compare \amulet with Nvidia DLSS Frame Generation, version 4.5~\cite{dlss4}, MobFGSR~\cite{yang2024mobfgsr} and MoFlow~\cite{yang2024mobfgsr}. 
\textcolor{red}{We could not include a comparison to GFFE~\cite{Wu2024}, because the code and training data for this method has not been made publicly available.}
DLSS is a professional tool for interpolation of up to five frames between ground truth images. MobFGSR is a state-of-the-art lightweight, non-neural frame extrapolation and interpolation method. MoFlow is a state-of-the-art frame extrapolation method based on neural networks. DLSS, MobFGSR and MoFlow were provided with conventional flat G-buffers (depth, color, alpha, motion vectors) from our deferred renderer and any additional inputs required.

We used reference code and models from public repositories. This results in some limitations. The public code sample of MobFGSR does not support frame extrapolation. Thus, we compared them with the frame interpolation method of MobFGSR. We could only source the single frame extrapolation model of MoFlow. Thus, we evaluated only the first generated frame of MoFlow. We did not attempt to retrain the MoFlow model, but we chose the model that best fits our scenes. This ensures a fair comparison with DLSS and MobFGSR, which were both used unmodified, and with \amulet which does not require training.

\amulet and DLSS Frame Generation run live in the Falcor engine, while MoFlow and MobFGSR work offline, generating frames using exported G-Buffer attributes. DLSS was configured to generate three novel frames. MobFGSR was configured to generate three novel frames from interpolating two ground truth images. MoFlow was configured to extrapolate one novel frame from a ground truth image. DLSS and MobFGSR can only perform frame interpolation, not frame extrapolation. Therefore, we compare its interpolated results with the ground truth in the same way as we do for the two frame extrapolation methods (\amulet and MoFlow). This decision grants DLSS and MobFGSR an advantage, as they are not penalized for the additional latency introduced by having to wait for the next keyframe to arrive. Additionally, interpolation methods know about objects moving in and out of the frustrum, which can create significant challenges for extrapolation methods. In summary, by comparing and competing with frame interpolation models, we put our approach at an artificial disadvantage. This procedure further strengthens our results.

\begin{table}[t]
\centering
\caption{Comparison of runtimes achieved by the baseline deferred renderer, \amulet, DLSS, MobFGSR. Amulet offers comparable runtime performance to DLSS and MobFGSR. }
\label{tab:runtimes}
\begin{tabular}{llrrrr}
\toprule
\textbf{Scene} & \textbf{Res.} & \textbf{Def.} & \textbf{Amulet} & \textbf{DLSS} & \textbf{MobFGSR} \\
\midrule
\multirow{2}{*}{\makecell[l]{Bistro\\Interior}}
& Full HD & 13.95 & 5.25 & 4.66 & \textbf{3.72}\\
& 4K      & 13.50 & \textbf{5.91} & 6.29 & 6.69\\
\midrule
\multirow{2}{*}{\makecell[l]{Bistro\\Exterior}}
& Full HD & 12.95 & 5.47 & 3.63 & \textbf{3.47}\\
& 4K      & 24.53 & 8.42 & \textbf{6.66} & 8.81\\
\midrule
\multirow{2}{*}{Sponza}
& Full HD & 15.28 & 5.00 & 4.28 & \textbf{4.02}\\
& 4K      & 15.70 & 6.87 & \textbf{5.49} & 7.06\\
\midrule
\multirow{2}{*}{\makecell[l]{San\\Miguel}}
& Full HD & 12.80 & 5.20 & 4.11 & \textbf{3.42}\\
& 4K      & 11.62 & 6.23 & \textbf{4.84} & 5.49\\
\bottomrule
\end{tabular}
\end{table}

\paragraph*{Performance}

\autoref{tab:runtimes} compares the runtimes of frame generation. For all methods, the shading load is divided by the period length (four frames), and the method-specific overhead is added (geometry and composition pass for \amulet and deferred rendering, composition pass for \amulet, neural network inference for DLSS and MoFlow).
DLSS and \amulet run at similar frame rates; however, interpolation makes DLSS late by one period (four frames). MoFlow is more than 4\ti\ slower in Full HD and could not support 4K at all on a GPU with 32\,GB. In Full HD, MoFlow allocates 9\,GB.
MoFlow takes approximately 17 ms to generate a frame, which is slower than the baseline deferred renderer.

\begin{table*}[t]
\centering
\caption{Image quality comparison between Amulet, DLSS, and MobFGSR. Higher PSNR and SSIM values indicate better quality, while lower LPIPS and FLIP values are preferred. Best results are highlighted in bold. Quality results were averaged over all scenes. In contrast to \amulet, DLSS and MobFGSR use interpolation, i.e., they have access to future information. \amulet yields comparable per-pixel error results (PSNR, SSIM) and exceeds MobFGSR and DLSS in perceptual error (FLIP, LPIPS).}
\label{tab:image_quality_comparison}
\resizebox{\textwidth}{!}{
\begin{tabular}{ll|rrrr|rrrr|rrrr|rrrr}
\toprule
& &
\multicolumn{4}{c|}{\textbf{Bistro Interior}} &
\multicolumn{4}{c|}{\textbf{Bistro Exterior}} &
\multicolumn{4}{c|}{\textbf{Sponza}} &
\multicolumn{4}{c}{\textbf{San Miguel}} \\
\cmidrule(lr){3-6}
\cmidrule(lr){7-10}
\cmidrule(lr){11-14}
\cmidrule(lr){15-18}
\textbf{Res.} & \textbf{Method} &
\textbf{PSNR}$\uparrow$ & \textbf{SSIM}$\uparrow$ & \textbf{LPIPS}$\downarrow$ & \textbf{FLIP}$\downarrow$ &
\textbf{PSNR}$\uparrow$ & \textbf{SSIM}$\uparrow$ & \textbf{LPIPS}$\downarrow$ & \textbf{FLIP}$\downarrow$ &
\textbf{PSNR}$\uparrow$ & \textbf{SSIM}$\uparrow$ & \textbf{LPIPS}$\downarrow$ & \textbf{FLIP}$\downarrow$ &
\textbf{PSNR}$\uparrow$ & \textbf{SSIM}$\uparrow$ & \textbf{LPIPS}$\downarrow$ & \textbf{FLIP}$\downarrow$ \\
\midrule
\multirow{3}{*}{Full HD}
& \amulet
& \textbf{29.52} & \textbf{0.9294} & \textbf{0.0175} & \textbf{0.0223}
& 27.65 & 0.9018 & \textbf{0.0292} & \textbf{0.0002}
& 32.27 & \textbf{0.9391} & \textbf{0.0142} & \textbf{0.0304}
& \textbf{31.11} & \textbf{0.9486} & \textbf{0.0157} & \textbf{0.0000} \\
& DLSS
& 23.92 & 0.8695 & 0.0628 & 0.0492
& \textbf{28.83} & \textbf{0.9193} & 0.0547 & 0.0002
& 30.55 & 0.9147 & 0.0394 & 0.0574
& 29.57 & 0.9267 & 0.0331 & 0.0000 \\
& MobFGSR
& 23.32 & 0.8517 & 0.0594 & 0.0518
& 28.43 & 0.9099 & 0.0336 & 0.0005
& \textbf{32.34} & 0.9362 & 0.0214 & 0.0561
& 30.68 & 0.9270 & 0.0182 & 0.0002 \\
\midrule
\multirow{3}{*}{4K}
& \amulet
& \textbf{30.69} & \textbf{0.9287} & \textbf{0.0199} & \textbf{0.0201}
& 28.96 & 0.9186 & \textbf{0.0264} & \textbf{0.0002}
& 27.98 & \textbf{0.9262} & 0.0325 & \textbf{0.0460}
& \textbf{32.92} & \textbf{0.9617} & \textbf{0.0121} & \textbf{0.0001} \\
& DLSS
& 23.87 & 0.8560 & 0.0671 & 0.0475
& \textbf{30.09} & \textbf{0.9315} & 0.0458 & 0.0002
& 29.56 & 0.9023 & 0.0402 & 0.0540
& 29.74 & 0.9307 & 0.0286 & 0.0001 \\
& MobFGSR
& 23.11 & 0.8265 & 0.0677 & 0.0512
& 28.60 & 0.9067 & 0.0305 & 0.0004
& \textbf{31.75} & 0.9253 & \textbf{0.0224} & 0.0520
& 30.39 & 0.9225 & 0.0193 & 0.0005 \\
\bottomrule
\end{tabular}
}
\end{table*}

\begin{table}[t]
\centering
\caption{Image quality comparison between \amulet and MoFlow on the first extrapolated frame. \amulet exceeds MoFlow in PSNR, SSIM and LPIPS. }
\label{tab:amulet_moflow_comparison}
\begin{tabular}{lrrrr}
\toprule
\textbf{Method} & \textbf{PSNR $\uparrow$} & \textbf{SSIM $\uparrow$} & \textbf{LPIPS $\downarrow$} & \textbf{FLIP $\downarrow$} \\
\midrule
\amulet & \textbf{30.726} & \textbf{0.944} & \textbf{0.016} & 0.012 \\
MoFlow  & 30.660 & 0.941 & 0.037 & \textbf{0.002} \\
\bottomrule
\end{tabular}
\end{table}

\paragraph*{Visual quality}

\autoref{tab:image_quality_comparison} compares the image quality achieved by all methods. \amulet consistently generates better image quality than DLSS and MobFGSR in Bistro Interior and San Miguel. Similarly, \amulet prevails in 9 out of 16 measurements in Bistro Exterior and Sponza, despite the advantage that DLSS and MobFGSR have due to their access to a future reference frame. ~\autoref{tab:amulet_moflow_comparison} compares the visual quality of the first extrapolated frame in \amulet and MoFlow. MoFlow beats \amulet using FLIP, but \amulet beats MoFlow using PSNR, SSIM and LPIPS. We conclude that the image quality of \amulet is, at least, on par with the latest neural frame extrapolation methods.

\begin{figure*}
    \centering
    \includegraphics[width=0.49\linewidth]{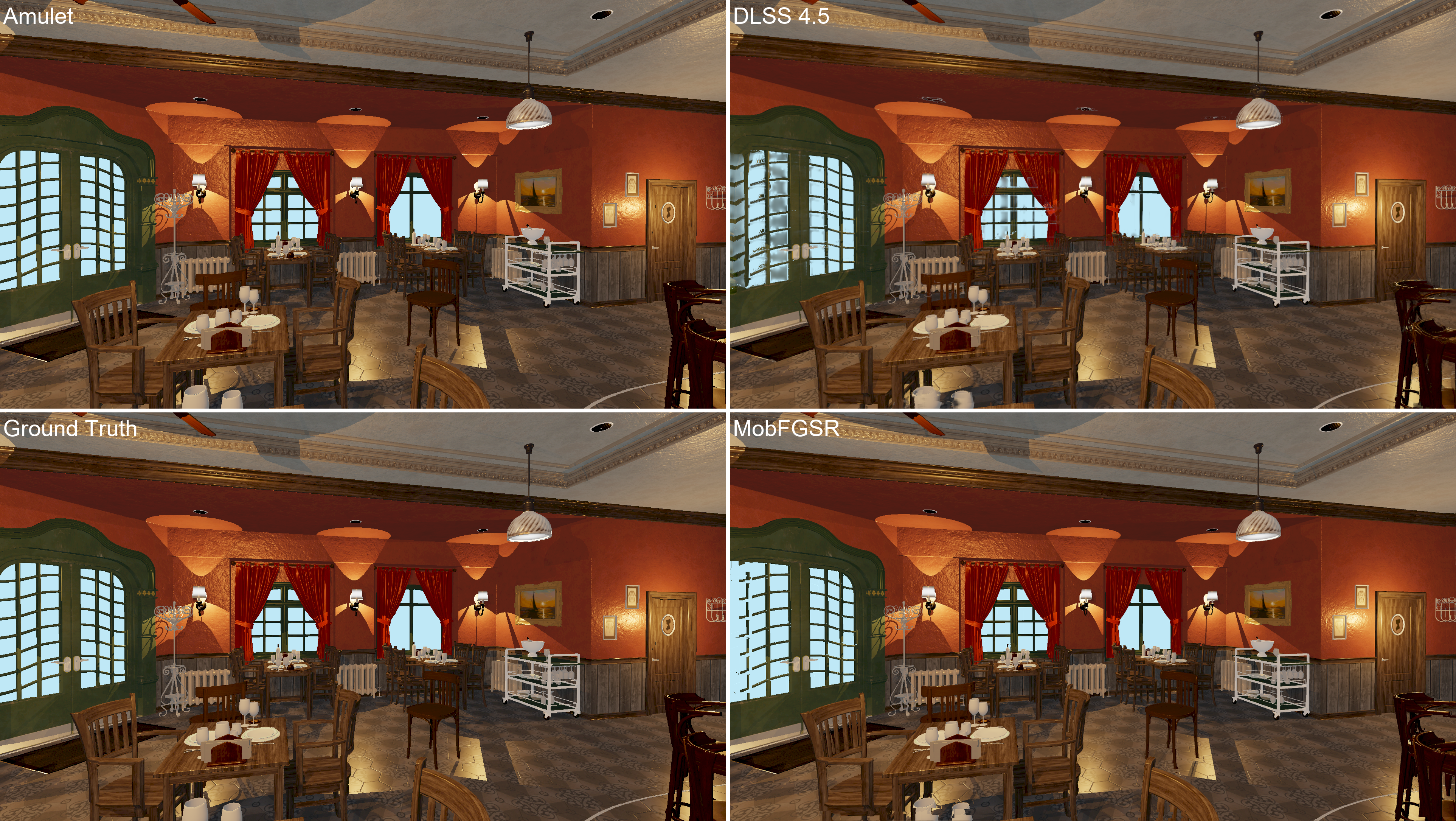}
    \includegraphics[width=0.49\linewidth]{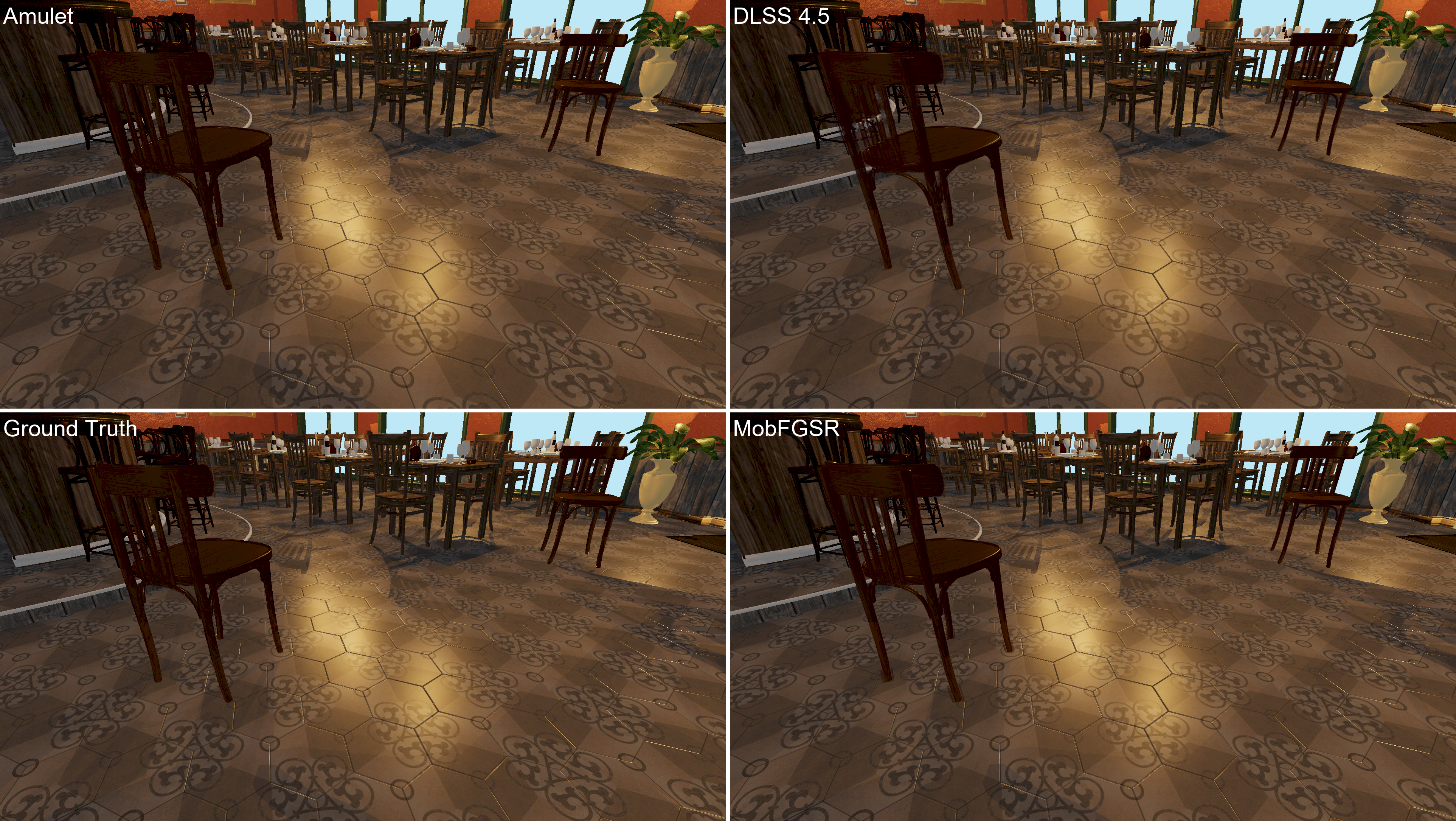}\\
    \includegraphics[width=0.49\linewidth]{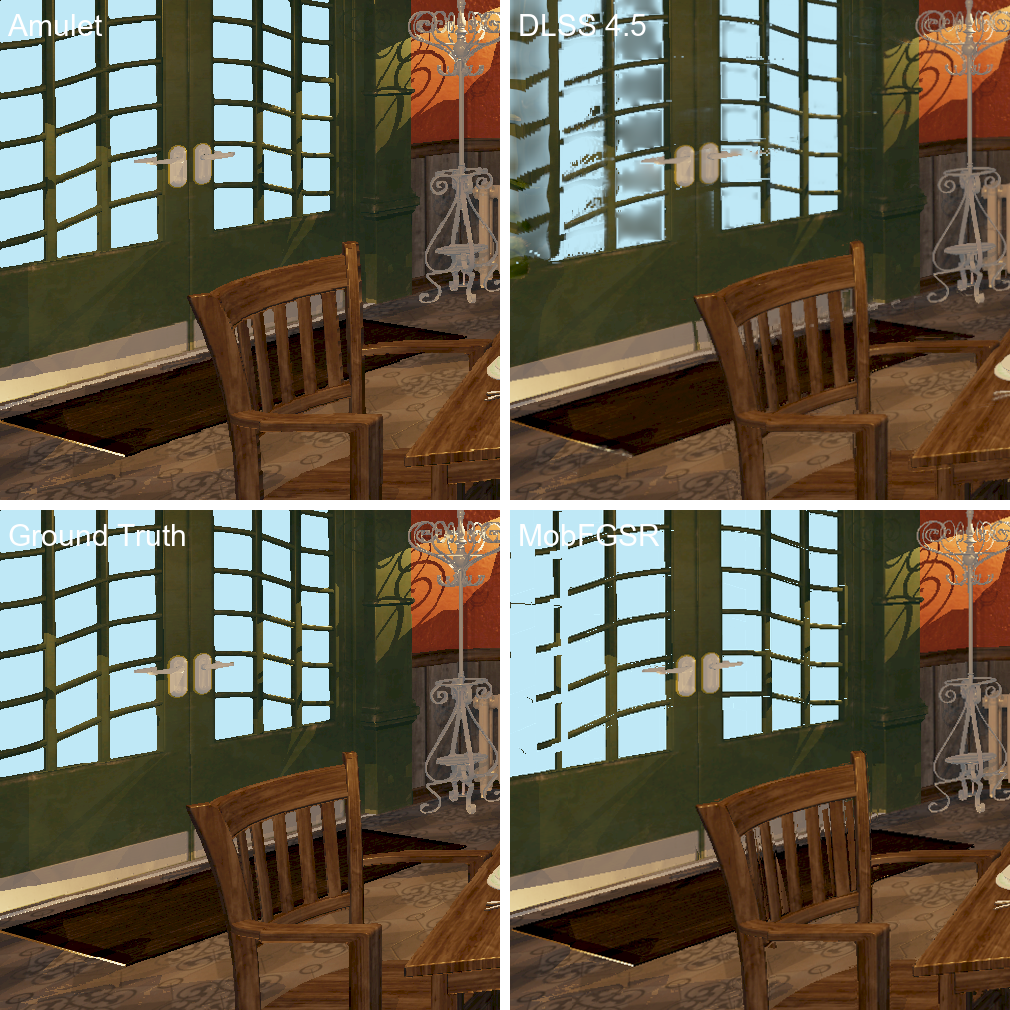}
    \includegraphics[width=0.49\linewidth]{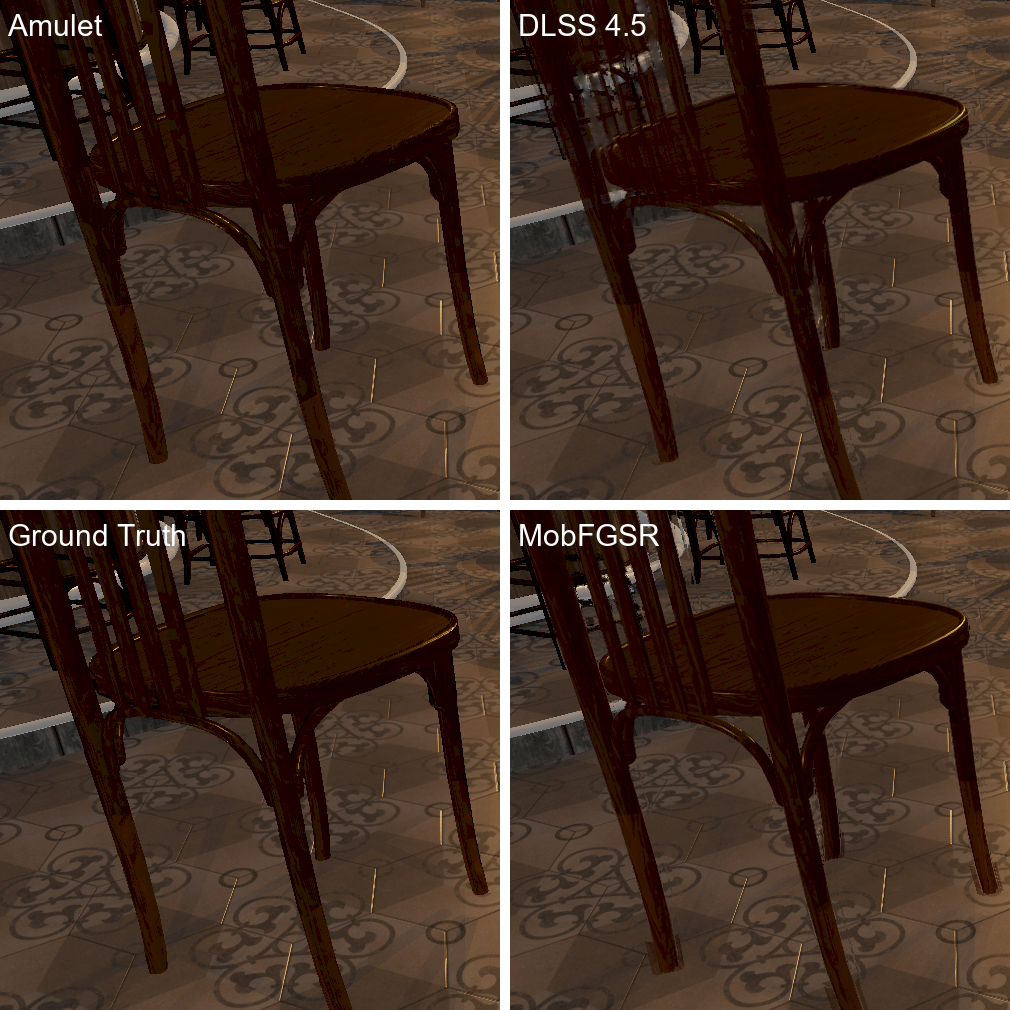}
    \caption{\textbf{Left}: Comparison of the first generated frame (after ground truth) between \amulet, DLSS 4.5 and MobFGSR. The ground truth is shown in the bottom left of the left column. In this view, we focus on \textit{artifacts from thin structures}, such as window frames. In the overview, we can observe additional artifacts on the wine glasses close to the camera for DLSS and MobFGSR. \amulet handles the shown structures well. \textbf{Right}: Comparison of the second generated frame (after ground truth) between \amulet, DLSS 4.5 and MobFGSR. The ground truth is shown in the bottom left of the left column. In this view, we focus on \textit{artifacts from disocclusion behind thin structures}, such as chair backs and chair legs. MobFGSR shows artifacts around the chair legs, DLSS around the chair back. \amulet does not show such artifacts. On the other hand, the highlight on the top right corner of the chair is handled well by DLSS and MobFGSR, but to small for the sub-sampled shading gradients of \amulet to pick up.}
    \label{fig:big_comparison_3}
\end{figure*}

\begin{figure*}
    \centering
    \includegraphics[width=0.49\linewidth]{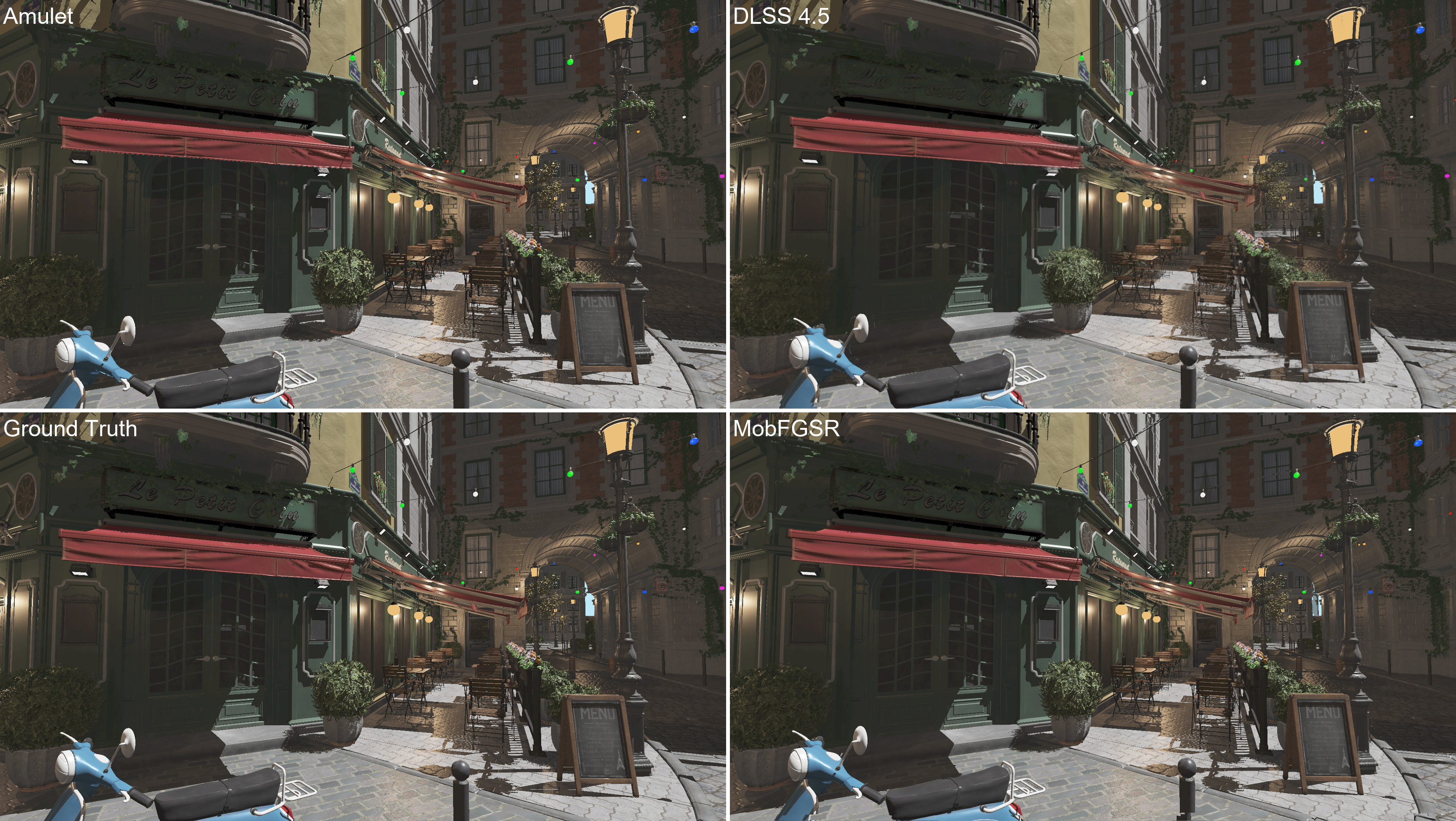}
    \includegraphics[width=0.49\linewidth]{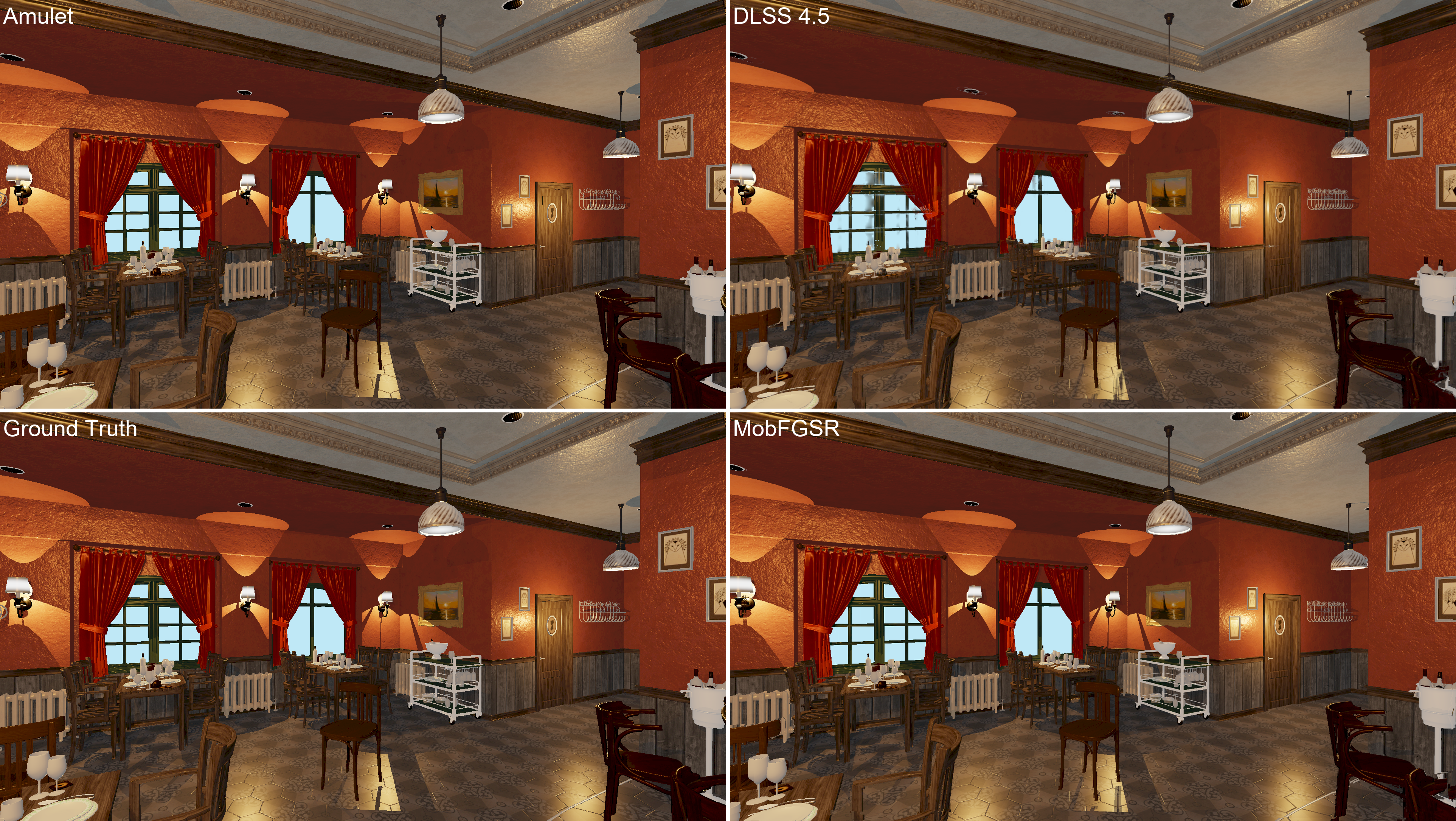}\\
    \includegraphics[width=0.49\linewidth]{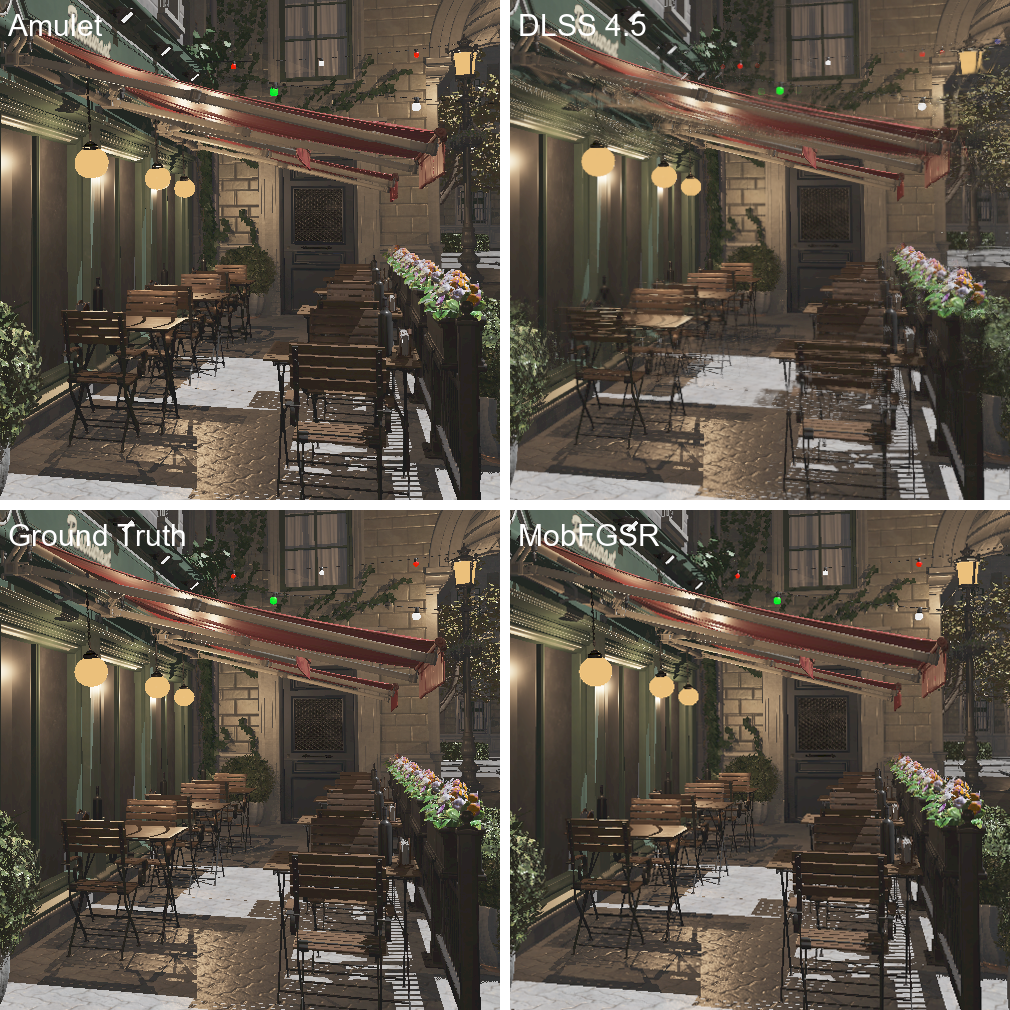}
    \includegraphics[width=0.49\linewidth]{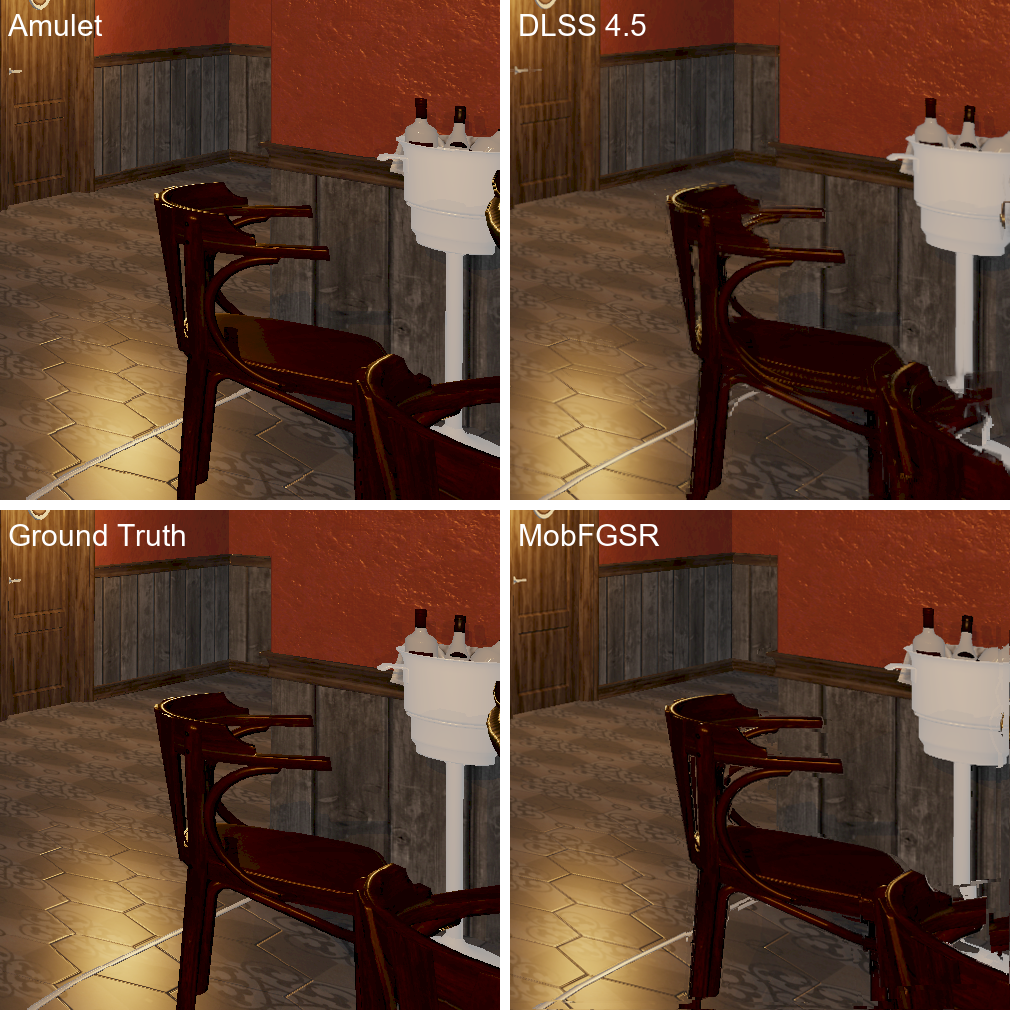}
    \caption{\textbf{Left}: Comparison of the second generated frame (after ground truth) between \amulet, DLSS 4.5 and MobFGSR. The ground truth is shown in the bottom left of the left column. In this view, we focus on \textit{blurring around detailed geometry}, such as garden chairs. \amulet and MobFGSR handle the detailed geometry adequately, while DLSS shows blurring around the chairs and around the mesh texture on the door in the back. \textbf{Right}: Comparison of the second generated frame (after ground truth) between \amulet, DLSS 4.5 and MobFGSR. We focus on \textit{objects entering the view from the right under rotating motion}. \amulet handles this scenario correctly by design, MobFGSR and DLSS suffer from warping, especially if geometry near the edge overlaps.}
    \label{fig:big_comparison_2}
\end{figure*}

\begin{figure*}
    \centering
    \includegraphics[width=0.49\linewidth]{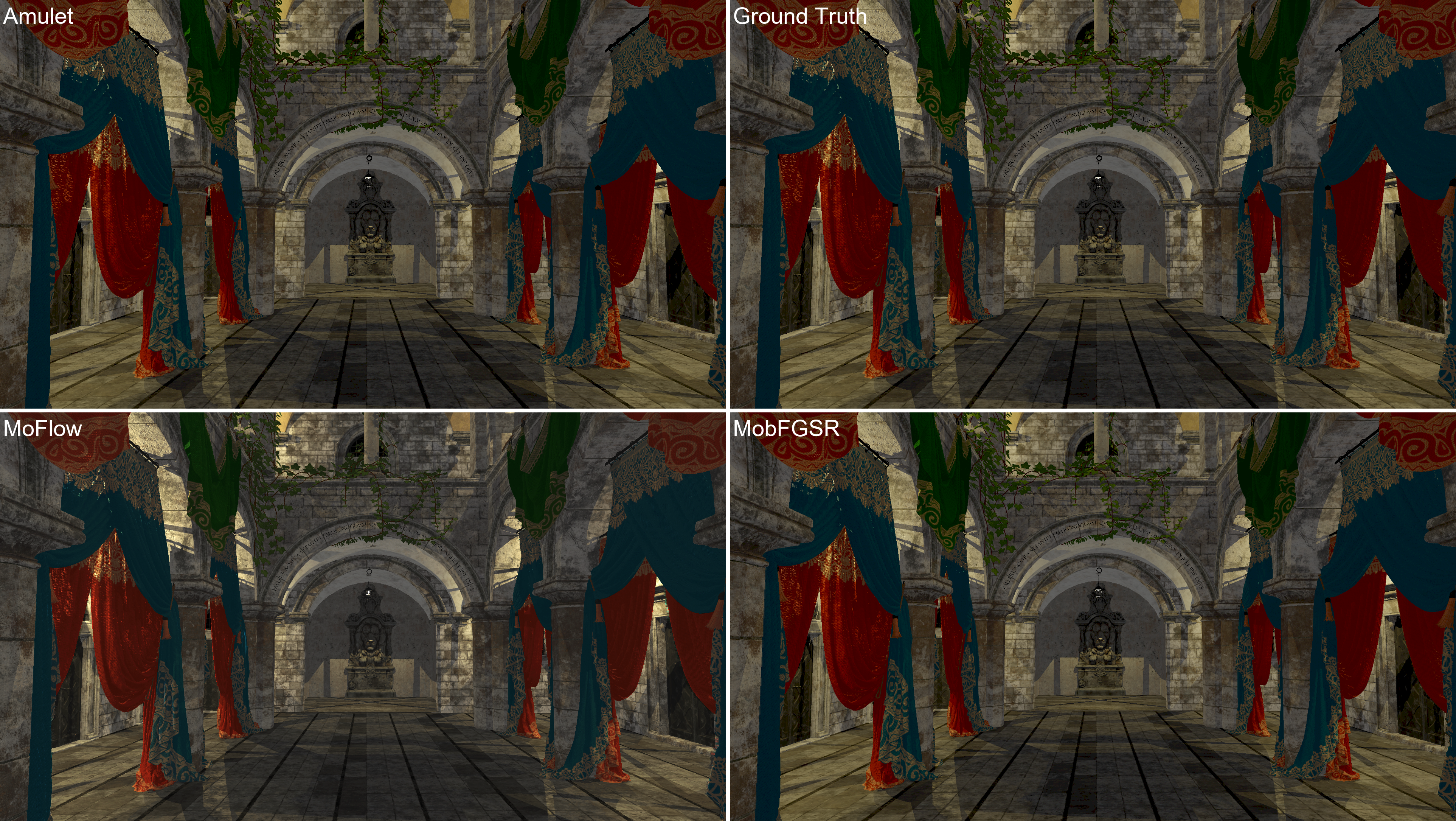}
    \includegraphics[width=0.49\linewidth]{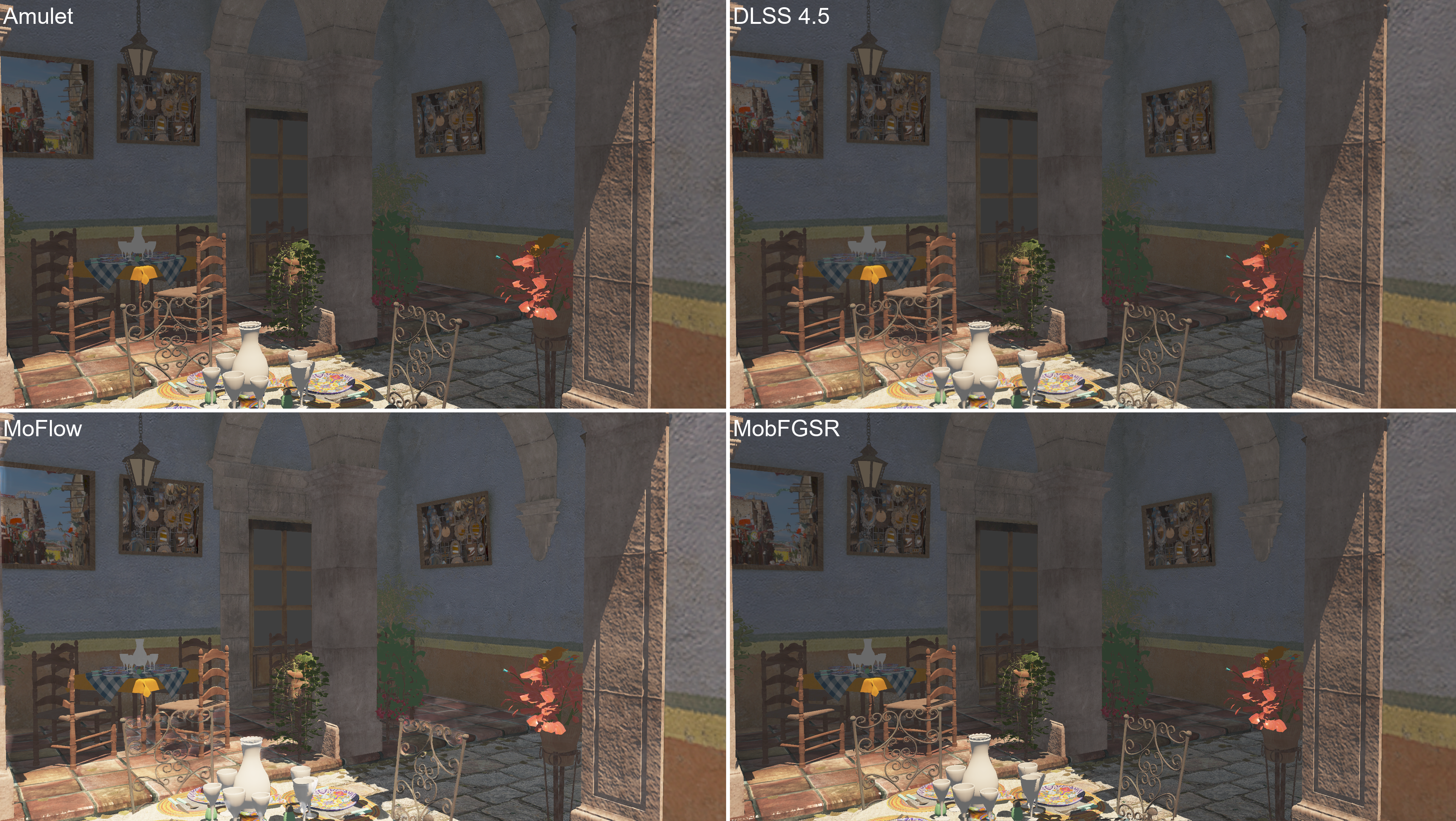}\\
    \includegraphics[width=0.49\linewidth]{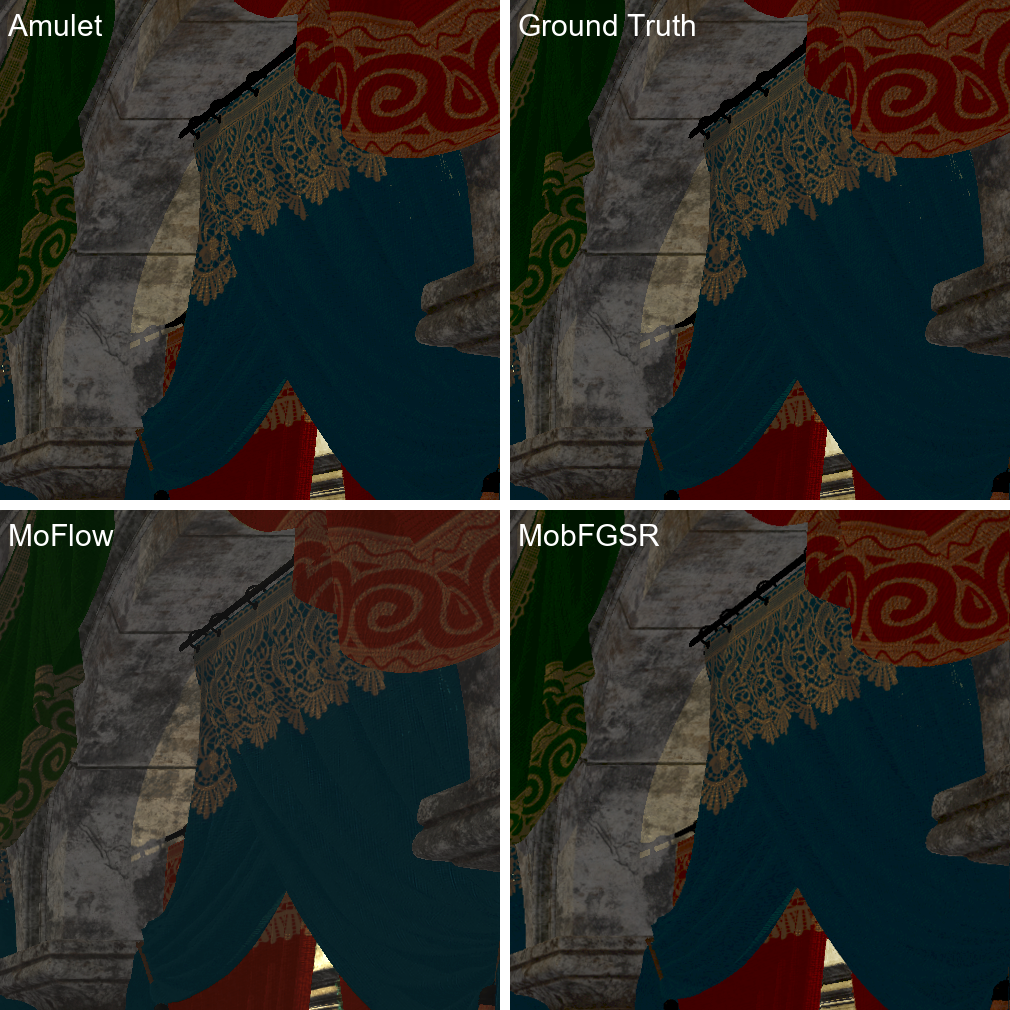}
    \includegraphics[width=0.49\linewidth]{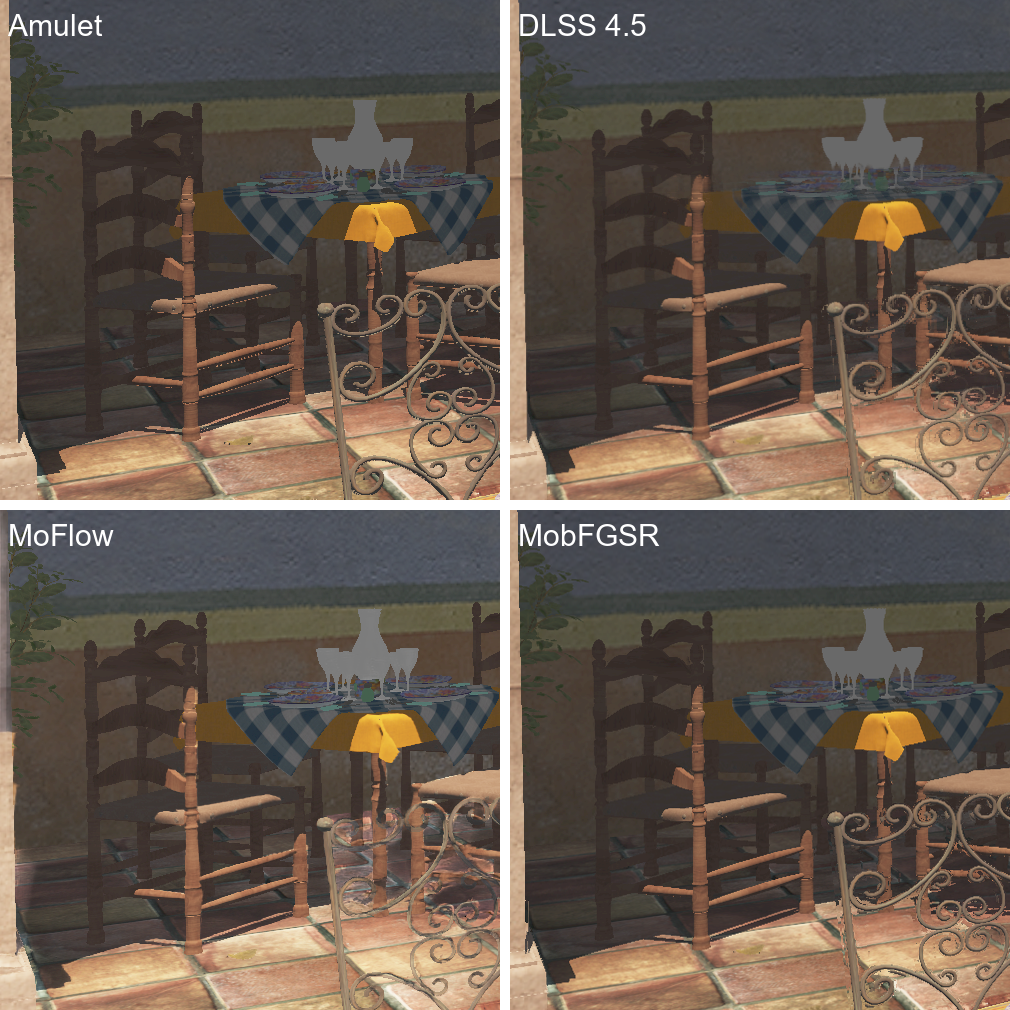}
    \caption{\textbf{Left}: Comparison of the first generated frame (after ground truth) between Amulet, MoFlow and MobFGSR. The ground truth is shown in the top right of the left column. In this view, all approaches yield strong extrapolation quality. \textbf{Right}: Comparison of the first generated frame (after ground truth) between Amulet, MoFlow and MobFGSR. In this view, there are two challenging regions. First, the column moving into view from the left under rotating motion. Amulet handles this naturally, DLSS and MobFGSR perform well. Interpolation methods are at an advantage in this scenario as the column entering is already known to them. MoFlow struggles to fill in the correct texturing. As the MoFlow model was not explicitly fine-tuned on this scene, this is unsurprising. Second, disocclusion artifacts around the wire chair occur in all approaches, with Amulet showing the best results and least visible artifacts}
    \label{fig:big_comparison_4}
\end{figure*}

\autoref{fig:big_comparison_3}, \ref{fig:big_comparison_2} and \ref{fig:big_comparison_4} compare the first or second extrapolated frames produced by \amulet with those produced by DLSS 4.5, MobFGSR, and the deferred renderer's ground truth. Both MobFGSR and DLSS can reconstruct the overall scene well, but introduce a number of visual artifacts. ~\autoref{fig:big_comparison_3} (left) shows artifacts around thin structures, such as window frames, and objects close to the camera, such as the frontmost wine glasses.  \amulet preserves the geometry in both cases, because the shading cache maintains high-frequency scene details.  

~\autoref{fig:big_comparison_3} shows blurry results around the moving chair in both MobFGSR and DLSS, where the colors of the chair and background blend due to missing scene information. \amulet can accurately recreate these scenarios due to its multi-layer view warping. ~\autoref{fig:big_comparison_2} contains examples similar to those in~\autoref{fig:big_comparison_3}, covering thin geometric structures in Bistro Exterior and objects that appear in the view frustum when the camera rotates. \autoref{fig:big_comparison_4} (left) compares \amulet with MoFlow and MobFGSR and demonstrates a scenario where all approaches yield high extrapolation quality. \autoref{fig:big_comparison_4} (right) compares all methods showcasing that only \amulet is capable of reconstructing the novel view without visual artifacts. \autoref{fig:big_comparison_1} explores another difficult frame generation scenario. A moving object (chair) casts a shadow on a textured surface. Frame generation must correctly transport the shadow along the surface without interfering with the surface texture. \amulet solves this issue by selectively reshading these regions with higher frequency. Shading gradients are used to reliably detect such regions.

\subsection{Longer prediction intervals}

\begin{figure}[t]
    \centering
    \includegraphics[height=2.9cm]{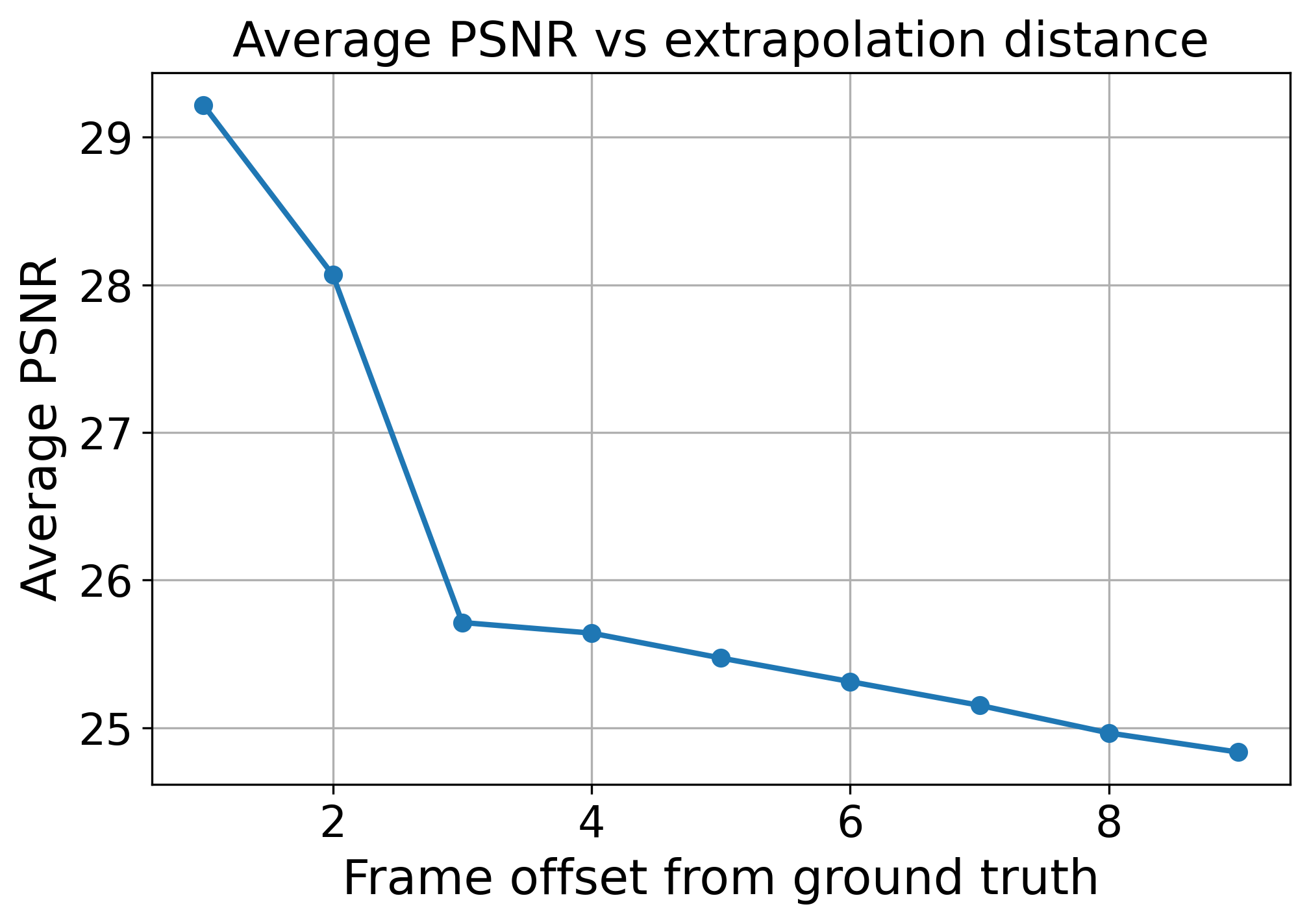}
    \includegraphics[height=2.9cm]{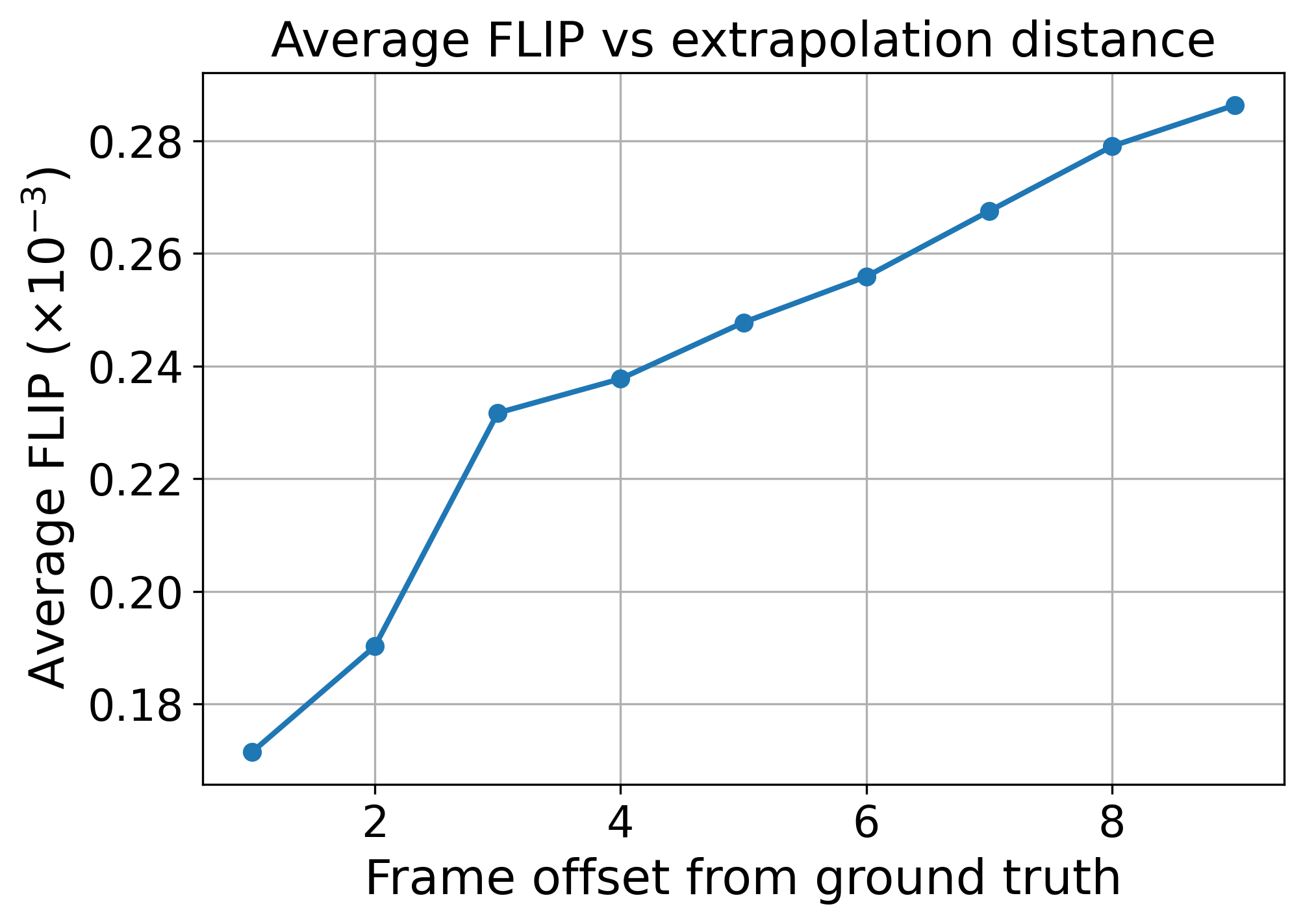}
    \caption{Visual error for a 9x extrapolation case in Bistro Exterior. The plot shows the \amulet PNSR and FLIP for the $n$-th frame after the ground truth. \amulet can produce high-quality results with minimal visual error, even for long prediction intervals. After the first and second frame, the quality of subsequent frames decreases linearly.}
    \label{fig:per_frame_errors}
\end{figure}

\autoref{fig:per_frame_errors}  shows the visual error for the $n$-th generated frame after the ground truth in \amulet, averaged over all frames in the Bistro Exterior dataset. In this scenario, we configured \amulet to produce nine frames after the ground truth. The first and second extrapolated frames are of the best quality. After that, the quality decreases linearly. This demonstrates \amulet's ability to produce high-quality results with minimal visual error and without noticeable artifacts over extended prediction intervals (\autoref{fig:teaser}).  \amulet can also extrapolate more than nine frames to support even more challenging scenarios. This flexibility is not provided by neural extrapolation methods that output a fixed number of frames.

\begin{figure*}
    \centering
    \includegraphics[width=0.99\linewidth]{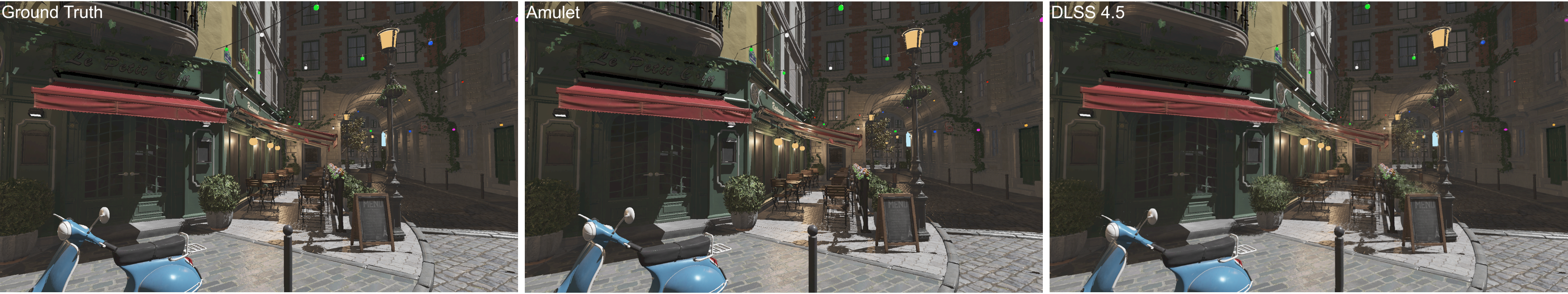}
    \caption{Comparison of \amulet vs. DLSS 4.5 6x multi-frame generation. Shown is the third generated frame after the ground truth by DLSS 4.5 (top), \amulet (middle) and the ground truth (bottom). DLSS 4.5 produces images with visual artifacts around detailed geometry (e.g. visible on the bistro chairs and the metal construction of the sunblind).}
    \label{fig:6x_visual_results}
\end{figure*}

\begin{table}[t]
\centering
\caption{Image quality comparison between \amulet and DLSS 4.5 averaged over the five generated frames. \amulet consistently outperforms DLSS 4.5 across all evaluated error metrics.}
\label{tab:6x_ablation}
\begin{tabular}{llrrrr}
\toprule
\textbf{Res.} & \textbf{Method} & \textbf{PSNR}$\uparrow$ & \textbf{SSIM}$\uparrow$ & \textbf{LPIPS}$\downarrow$ & \textbf{FLIP}$\downarrow$ \\
\midrule
\multirow{2}{*}{Full HD}
& \amulet & \textbf{25.61} & \textbf{0.8582} & \textbf{0.0428} & \textbf{0.0026} \\
& DLSS   & 23.38 & 0.7404 & 0.0907 & 0.0041 \\
\midrule
\multirow{2}{*}{4K}
& \amulet & \textbf{27.19} & \textbf{0.8874} & \textbf{0.0380} & \textbf{0.0020} \\
& DLSS   & 22.86 & 0.7226 & 0.0867 & 0.0039 \\
\bottomrule
\end{tabular}
\end{table}

The current state of the art method that outputs more than three frames is DLSS 4.5 which supports a 6$\times$ mode for dynamic multi-frame generation. This means that DLSS can generate five frames from ground truth. Note that DLSS performs frame interpolation rather than extrapolation. In other words, it uses past and future frames to generate intermediate frames, which gives it an advantage over \amulet, which does not have access to future information. \autoref{tab:6x_ablation} shows visual error results for DLSS 4.5 and \amulet when predicting five frames for a camera sequence in Bistro Exterior. \amulet outperforms DLSS 4.5 in this scenario in all error metrics. While \amulet is able to produce consistent frames from the layered cache, DLSS 4.5 suffers from hallucinations, ghosting artifacts and blurry results. This effect is amplified when the view rotates. An example of visual results is shown in~\autoref{fig:6x_visual_results}. In this example, the results for the chairs, tables and the balcony blinds are blurry, and the Vespa and billboard suffer from ghosting artifacts. Due to its non-neural nature, \amulet does not suffer from clearly visible artifacts even in challenging scenarios like this. In general, the number of artifacts that DLSS 4.5 produces increases with the distance a generated frame has from one of its anchor frames. 

\subsection{Shading gradients}

\begin{figure}
    \centering
    \includegraphics[width=0.99\linewidth]{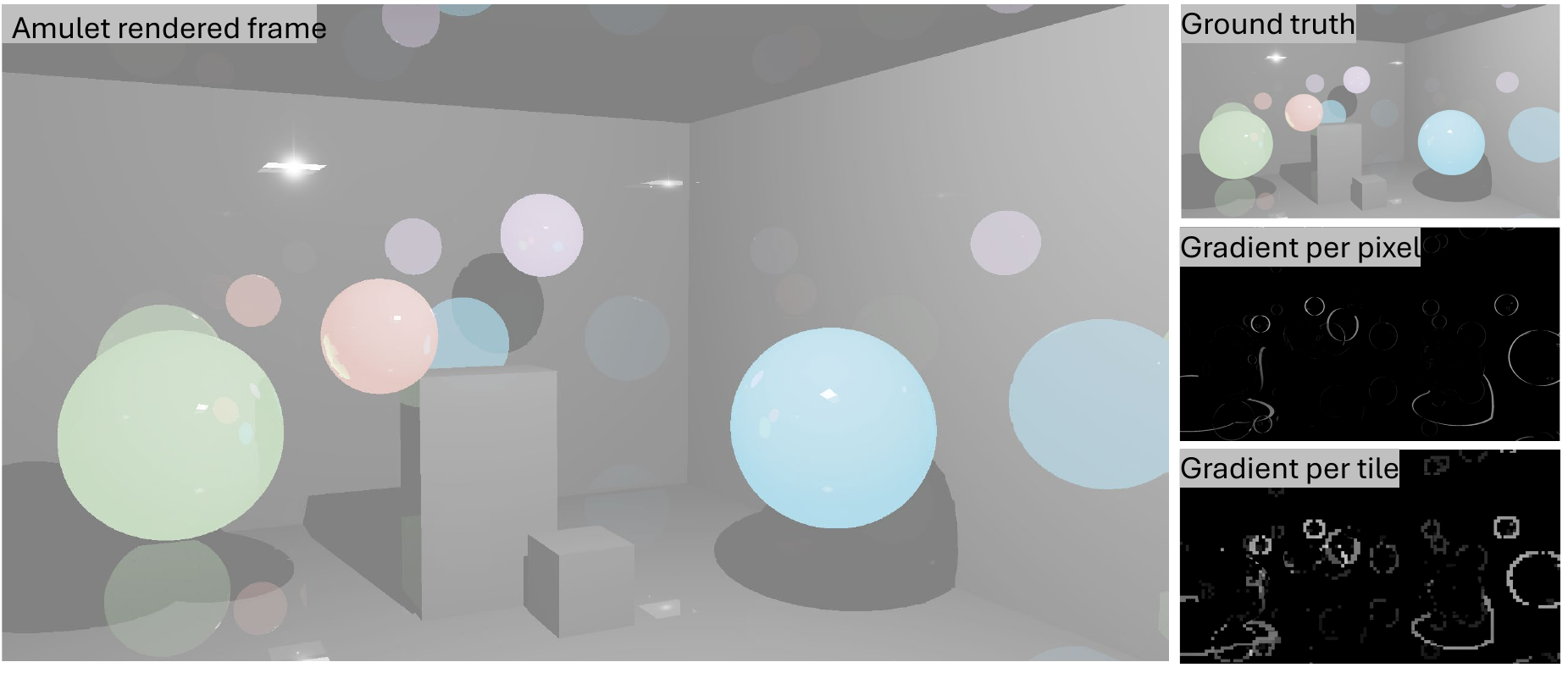}
    \caption{A mirror room with emissive moving objects and multi bounce specular reflection. Shading gradients adequately capture the movement of objects and reflections and allow selective updating of the shading cache in regions of high change (gradient magnitude). The per tile gradients are spread to neighbors with an additional max-convolution (not shown).}
    \label{fig:mirror_room}
\end{figure}

\begin{table}[t]
\centering
\caption{Image quality comparison between \amulet, DLSS, and MobFGSR for a scene with a moving object featuring moving shadows and specular highlights. Each moving object is placed in its own subcache, and shading gradients are enabled. Each method predicts three frames after the ground truth. \amulet achieves higher reconstruction quality by updating shading for challenging regions on the fly.}
\label{tab:chair_ablation}
\begin{tabular}{lrrrr}
\toprule
\textbf{Method} & \textbf{PSNR $\uparrow$} & \textbf{SSIM $\uparrow$} & \textbf{LPIPS $\downarrow$} & \textbf{FLIP $\downarrow$} \\
\midrule
\amulet & \textbf{40.706} & \textbf{0.991} & \textbf{0.003} & \textbf{0.014} \\
DLSS    & 24.563 & 0.928 & 0.049 & 0.070 \\
MobFGSR & 24.516 & 0.930 & 0.048 & 0.070 \\
\bottomrule
\end{tabular}
\end{table}

\begin{figure*}
    \centering
    \includegraphics[width=0.8\linewidth]{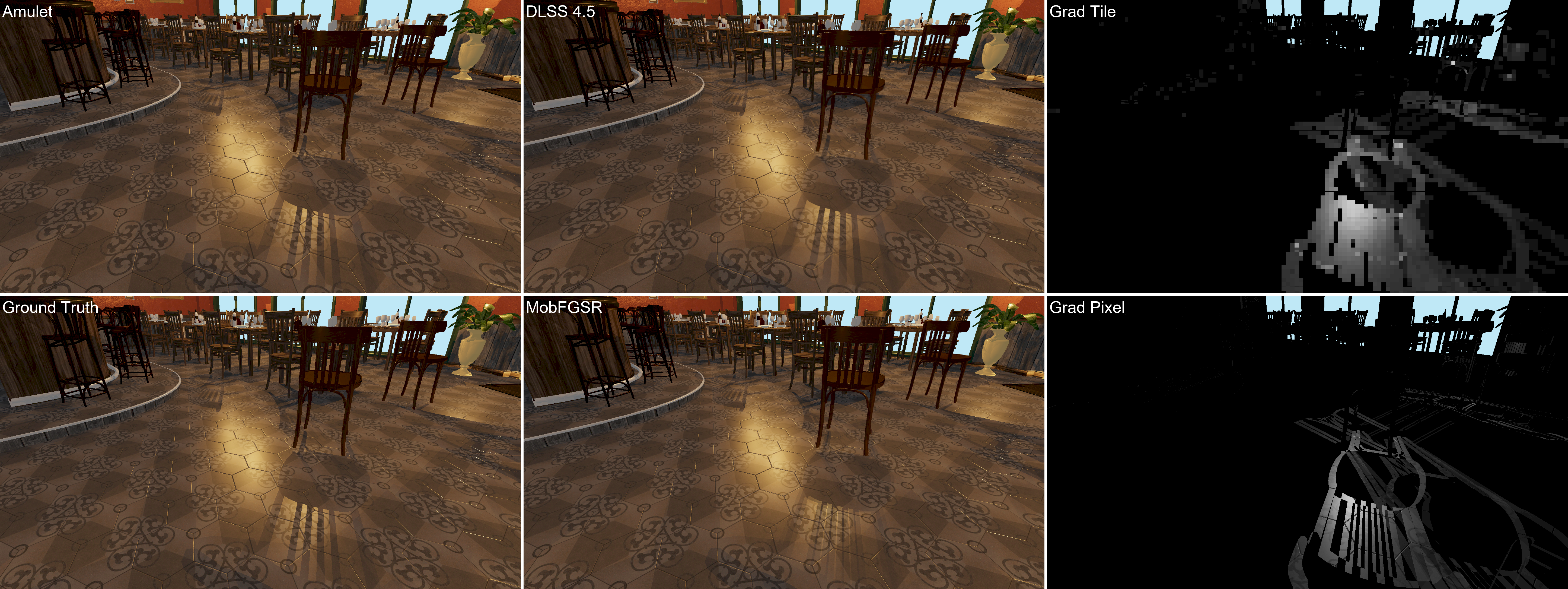}\\
    \includegraphics[width=0.8\linewidth]{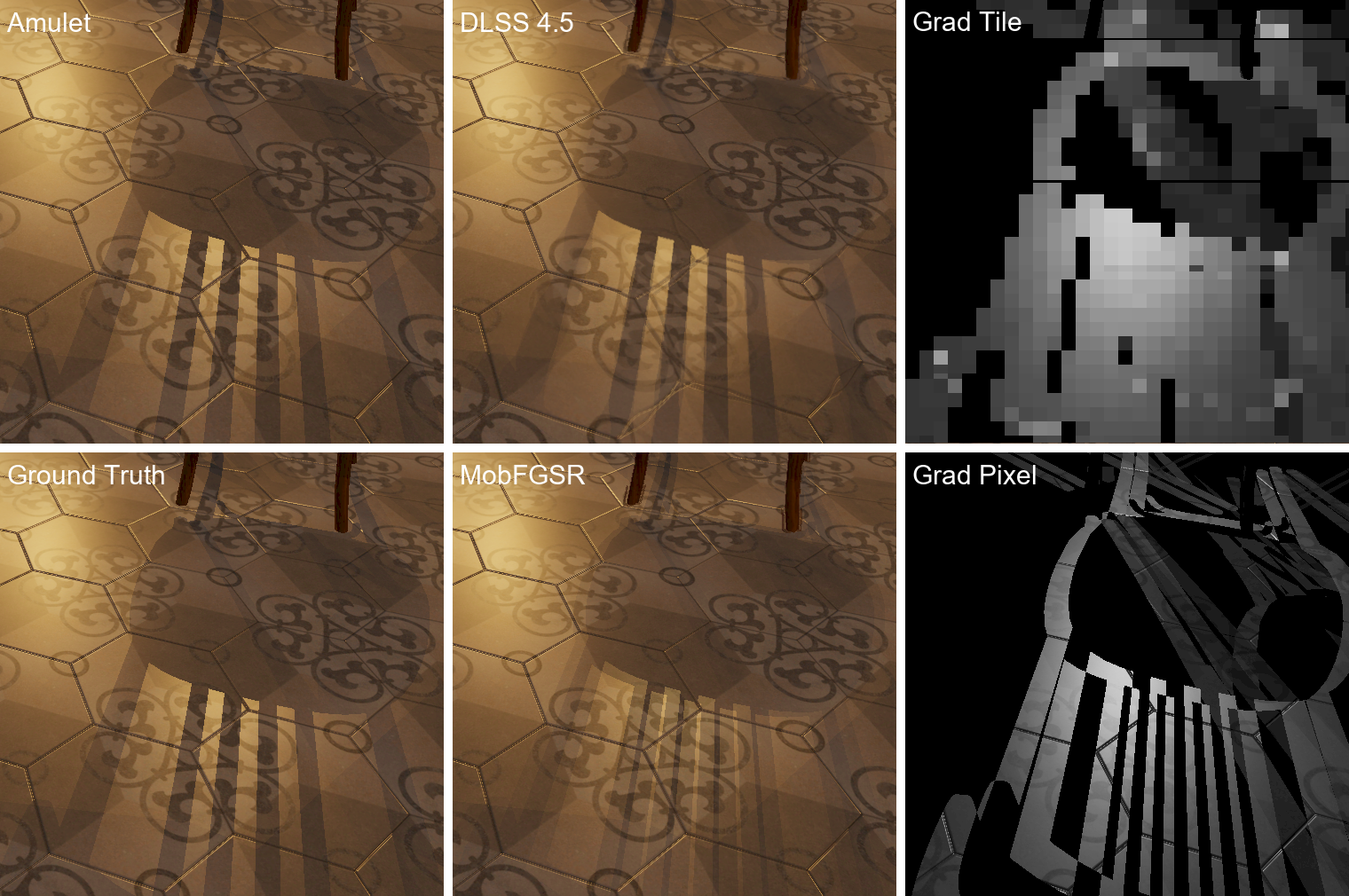}
    \caption{Two columns on the left: Comparison of the second generated frame (after ground truth) between \amulet, DLSS 4.5 and MobFGSR. The ground truth is shown in the bottom left of the left column. In this view, we focus on the shadow of a moving object on a textured surface. MobFGSR interpolated using motion vectors and depth. It cannot transport moving shadows and blends the anchor frames. DLSS transports the shadows mostly correct, but smears the floor texture of similar brightness. \amulet correctly reshades the moving shadow using shading gradients (right column). Shading gradients detect the regions of shading change and are used by \amulet to selectively update tiles at higher frequency.}
    \label{fig:big_comparison_1}
\end{figure*}

Shading gradients are designed to update shading more frequently in challenging scenarios. These scenarios include direct lighting effects from moving objects, such as shadow borders, and indirect lighting effects, such as reflections of moving objects. Our shading gradients support all of these scenarios and guarantee instant reshading of tiles with rapidly changing shading information between cache swaps.
~\autoref{fig:big_comparison_1} shows a moving and rotating chair in Bistro Interior. The chairs cast shadows on a textured surface. Frame generation must correctly transport the shadows along the surface without interfering with the surface texture. We measured the image quality in this scene for all methods in~\autoref{tab:chair_ablation}. \amulet  significantly outperforms both DLSS and MobFGSR, demonstrating its ability to detect frequent shading changes, such as those caused by the chair rods, and update all affected tiles in real time. 

Furthermore, shading gradients can even handle scenes with complex lighting effects, such as moving reflections. To stress-test our approach, we designed a synthetic scene with four animated, reflective, and emissive spheres inside a room with reflective walls, floor, and ceiling. We use raytracing with up to three bounces per ray to calculate the shading. The result is a scene with many shadows, reflections, and inter-reflections. See~\autoref{fig:mirror_room} for a visualization of the per-pixel and per-tile shading gradients. Tiles are frequently reshaded if the gradient exceeds a threshold. \amulet correctly recognizes moving objects, moving shadow borders on the floor, and moving reflections, such as sphere reflections on the wall, ceiling, and other spheres. Overall, \amulet can refresh outdated shading information to prevent stuttering lighting effects even over longer prediction intervals.

\subsection{Limitations}

The primary limitation of Amulet is highly dynamic scenes involving non-rigid transformations, such as deformations or animated characters. Such cases cannot be handled purely through subcache transformations and instead fall back to adaptive shading. Consequently, rendering cost increases with the screen-space area occupied by such dynamic content. In the worst case, every tile may require per-frame updates. An example is a nightclub scene with many animated characters and rapidly changing lighting. In such scenarios, fallback to deferred rendering, alternative extrapolation methods, or hybrid integration of Amulet with motion-vector-based approaches becomes reasonable. In summary, all current frame inter- and extrapolation methods have trade-offs. While Amulet addresses many existing limitations, handling highly dynamic scenes will require additional research.

\section{Conclusions}
\label{chap:conclusion}

We have introduced \amulet, an efficient frame extrapolation method that predicts novel views without relying on neural networks. A tiled, sparse, and layered cache stores samples in separate image layers, allowing accurate extrapolation in most disocclusion situations. Accelerated traversal of the cache along view rays also allows us to handle the contribution of semitransparent objects. Moreover, we detect dynamic shading changes that appear on visible samples by analyzing shading gradients across frames. We re-shade areas with rapidly changing shading on demand. Our comparisons with state-of-the-art methods reveal that our method generates (in most cases) higher quality at equivalent speed. 

While our method is more invasive to existing rendering frameworks than simply applying image filters or neural networks in frame postprocessing or in the driver, \amulet demonstrates the potential to scale rendering throughput to levels that have previously been unattainable. This potential is demonstrated even without using any spatial upscaling, a technique that can be considered orthogonal to frame extrapolation. We believe that graphics pipelines such as that of \amulet can finally depart from a philosophy of operating within the constraints of a fixed-function GPU pipeline. 

\begin{acks}
This work was supported by by the Alexander von Humboldt Foundation funded by the German Federal Ministry of Research, Technology and Space, German Research Foundation DFG (495135767), 
German Research Foundation \emph{DFG} (grant 528364066), German Research Foundation \emph{DFG} EXC 2120/2 - 390831618.
\end{acks}

\newpage
\bibliographystyle{ACM-Reference-Format}
\bibliography{references}

\newpage

\end{document}